# Mechanisms of Zero-Lag Synchronization in Cortical Motifs

Leonardo L. Gollo[1*], Claudio Mirasso[2], Olaf Sporns[3], Michael Breakspear[1,4,5]

[1] Systems Neuroscience Group, Queensland Institute of Medical Research, Brisbane, Queensland, Australia,

[2] IFISC, Instituto de Física Interdisciplinar y Sistemas Complejos (CSIC-UIB), Campus Universitat de les Illes Balears, E-07122, Palma de Mallorca, Spain,

[3] Department of Psychological and Brain Sciences, Indiana University, Bloomington, Indiana, United States of America

[4] School of Psychiatry, University of New South Wales and The Black Dog Institute, Sydney, New South Wales, Australia,

[5] The Royal Brisbane and Woman's Hospital, Brisbane, Queensland, Australia,

*Corresponding author email address: leonardo.l.gollo@gmail.com

## Abstract

Zero-lag synchronization between distant cortical areas has been observed in a diversity of experimental data sets and between many different regions of the brain. Several computational mechanisms have been proposed to account for such *isochronous synchronization* in the presence of long conduction delays: Of these, the phenomenon of "dynamical relaying" - a mechanism that relies on a specific network motif - has proven to be the most robust with respect to parameter mismatch and system noise. Surprisingly, despite a contrary belief in the community, the common driving motif is an unreliable means of establishing zero-lag synchrony. Although dynamical relaying has been validated in empirical and computational studies, the deeper dynamical mechanisms and comparison to dynamics on other motifs is lacking. By systematically comparing synchronization on a variety of small motifs, we establish that the presence of a single reciprocally connected pair - a "resonance pair" - plays a crucial role in disambiguating those motifs that foster zero-lag synchrony in the presence of conduction delays (such as dynamical relaying) from those that do not (such as the common driving triad). Remarkably, minor structural changes to the common driving motif that incorporate a reciprocal pair recover robust zero-lag synchrony. The findings are observed in computational models of spiking neurons, populations of spiking neurons and neural mass models, and arise whether the oscillatory systems are periodic, chaotic, noise-free or driven by stochastic inputs. The influence of the resonance pair is also robust to parameter mismatch and asymmetrical time delays amongst the elements of the motif. We call this manner of facilitating zero-lag synchrony *resonance-induced synchronization,* outline the conditions for its occurrence, and propose that it may be a general mechanism to promote zero-lag synchrony in the brain.

## Author Summary

Understanding large-scale neuronal dynamics - and how they relate to the cortical anatomy - is one of the key areas of neuroscience research. Despite a wealth of recent research, the key principles of this relationship have yet to be established. Here we employ computational modeling to study neuronal dynamics on small subgraphs - or motifs - across a hierarchy of spatial scales. We establish a novel organizing principle that we term a "resonance pair" (two mutually coupled nodes), which promotes stable, zero-lag synchrony amongst motif nodes. The bidirectional coupling between a resonance pair acts to mutually adjust their dynamics onto a common and relatively stable synchronized regime, which then propagates and stabilizes the synchronization of other nodes within the motif. Remarkably, we find that this effect can propagate along chains of coupled nodes and hence holds the potential to promote stable zero-lag synchrony in larger sub-networks of cortical systems. Our findings hence suggest a potential unifying account of the existence of zero-lag synchrony, an important phenomenon that may underlie crucial cognitive processes in the brain. Moreover, such pairs of mutually coupled oscillators are found in a wide variety of physical and biological systems suggesting a new, broadly relevant and unifying principle.

## Introduction

The study of large-scale brain dynamics, and the cortical networks on which they unfold, is a very active research area, providing new insights into the mechanisms of functional integration and complementing the traditional focus on functional specialization in the brain [1,2]. Whilst progress towards understanding the underlying network structure has been impressive [3,4], the emergent network dynamics and the constraints exerted on these dynamics by the network structure remain poorly understood [5]. The problem is certainly not straightforward, as the dynamics between just a pair of neural regions already depends critically on the nature of the local dynamics and the nature of the coupling between them [6]: Although non-trivial, a complete description of nonlinear dynamics between a pair of nodes is nonetheless typically possible [7]. However, aggregating such duplets into larger arrays and introducing noise and time delays leads to further challenges and prohibits an exact description of the precise functional repertoire, motivating recourse to the broader objective of finding unifying and simplifying principles [8]. Structural and functional *motifs* - small subnetworks of larger complex systems - represent such a principle [9]. As depicted in Fig. 1 a, they characterise an intermediate scale of organization between individual nodes and large-scale networks that may play a crucial role as elementary building blocks of many biological systems [10]. Motif distribution in cortical networks has also been shown to be highly non-random, with a small set of motifs that appear to be significantly enriched in brain networks [9]. The relative occurrence of 3-node motifs in three different anatomical networks of the Macaque brain and cat cortex (Fig. 1 b-e) is shown in Fig. 1 f-i. These motifs may play distinct roles in supporting various computational processes. In this report we examine the principles of neuronal dynamics that emerge on small motifs and consider their putative role in neuronal function.

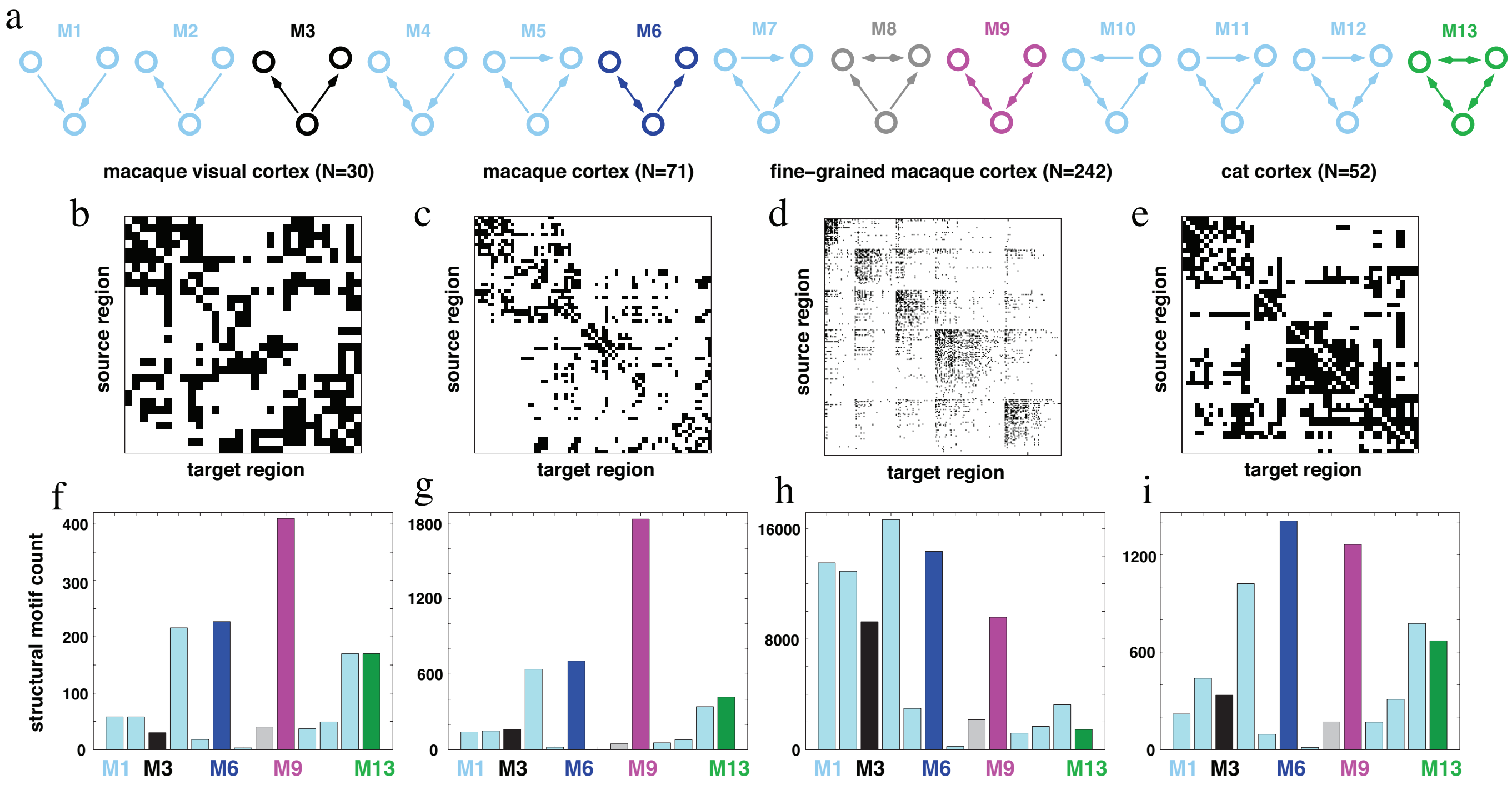

*Figure 1: Motifs in cortical networks. (a) The thirteen different motifs of size 3. (b-e) Connectivity matrices, and (f-i) Structural motif counts for each cortical network. Data (from the CoCoMac database [67, 68]) and algorithms are available at the brain connectivity toolbox website [69].*

The mechanisms supporting zero-lag synchrony between spatially remote cortical regions can be considered paradigmatic of those mediating between structure and function. Since first reported in cat visual cortex [11], zero-lag synchrony has been widely documented in empirical data and ascribed a range of crucial neuronal functions, from perceptual integration to the execution of coordinated motor behaviours [12-16]. In particular, zero-lag synchrony between populations of neurons (quantified through synchrony between the local field potentials) may play a crucial role in aligning packets of spikes into critical windows to maximise the reliability of information transmission at the neuronal level [17], and to bring mis-aligned spikes into the time window of spike-time-dependent plasticity [18]. The situation is particularly pertinent in sensory systems, where precise differences in the timing of inputs, between left and right cortex for example, may carry crucial information about the spatial location of the perceptual source [19]. However, the empirical occurrence of zero-lag synchronization is at apparent odds with the observation that two mutually–coupled oscillators interacting through a time-delayed connection do not, in general, exhibit zero-lag synchrony [20]. Indeed, in many models of neuronal systems the presence of a reciprocal delay has been found to introduce a 'frustration' into the system such that zero-lag synchrony is unstable and out-of-phase synchrony is instead the preferred dynamic relationship [21]. In fact, this phenomenon occurs quite generally in systems of oscillators with time-delayed coupling [21,22].

Complex dynamics in spatially embedded systems arise in a broad variety of physical and biological contexts. Arrays of coupled semiconductor lasers are a prominent example. Because of their extraordinary internal speed, even small time delays due to the finite speed of light are usually non-negligible in arrays of coupled lasers [23]. Detailed analysis of delay-coupled laser systems has suggested that an intermediate and reciprocally coupled relay node in a motif of three nodes could represent a general mechanism for promoting zero-lag synchrony in delay-coupled systems [24]. In previous work, it was also shown that such motif arrangements also represent a candidate mechanism for zero-lag synchrony in delay-coupled neuronal systems [25]. This is encouraging because there exist several candidate neuronal circuits in the mammalian brain which are characterized by reciprocal coupling between an intermediate delay node, including corticothalamic loops and the hippocampus [26,27]. There also exist strong reciprocal connections in the visual system, such as the heavily myelinated connections between primary visual cortex and the frontal eye fields. Indeed, the corresponding motif occurs disproportionally in mammalian cortex (Fig. 1), hence being embedded in many cortical subsystems [9].

The presence of a node that drives two common-driven nodes that reach zero-lag synchrony between them due to the driver's influence is intuitively appealing and finds anatomical support, for example, by shared input through bifurcating axons [13]. Certainly, a common-driving input of sufficient intensity can generate virtually perfect spike-time correlation, as long as the time delay to both driven nodes is identical. However, this scenario is not robust if the time delays lose symmetry or the coupling is not sufficiently strong. The common-driving setup is, however, a key prototype that offers insights into the synchronization between the driven nodes and the roles of the dynamics of the nodes [28-32]. Here we consider dynamics on the 3-node motifs that occur abundantly in large-scale networks of the brain (Fig. 1), adding connections to the prototypical common-driving motif. We confirm that common driving - a coupling arrangement that is widely invoked in the literature - is an ineffective means of inducing zero-lag synchrony in the presence of weak coupling (a neurophysiologically plausible regime). However, the additional incorporation of a single reciprocally coupled connection between the driver and an edge node - which leads to synchrony between that pair - is found to be a novel and efficient way of promoting zero-lag synchrony amongst other nodes in these small motifs. We demonstrate that this effect - which we term *resonance-induced synchrony* – arises consistently in candidate computational models at the neuronal, population and mesoscopic spatial scales and is robust to mismatches in system parameters and even time delays. Remarkably, we show that the resonance effects of a synchronized pair are not necessarily localized, but may instead propagate throughout the network. We hence propose resonance-induced synchrony as a general and unifying mechanism of facilitating zero-lag synchrony in the brain.

## Results

We studied zero-lag synchronization - quantified as the average zero-lag cross-correlation between two nodes A and B ($C_{AB}$) - in a variety of different motifs involving a common driving node. We considered the dynamics of nodes expressing different neuronal systems across a hierarchy of scales. At the microscopic scale, each node was modelled to represent a single spiking Hodgkin-Huxley neuron; at the circuit scale, each node was taken to represent a population of 400 excitatory and 100 inhibitory randomly connected neurons described by the Izhikevich model; and at the mesoscopic scale each node was modelled as a neural mass model with chaotic activity. This last model permits systematic parameter exploration that is not possible with populations of spiking neurons. In all the three modeling levels, coupling between nodes was via excitatory chemical synapses (see Methods for details on models and integration scheme). For the sake of simplicity, we initially assumed homogeneous delays in the motifs, i.e., all connections between nodes had the same time delay. We later explored the robustness of the results when relaxing these assumptions in the section "Mismatch in the conduction delays".

The notation we adopt for the motifs of three nodes follows the notation of Sporns and Kötter (2004)

[9] who denoted all 13 possible connected subgraphs (motifs) composed of three nodes, denoted from M1 to M13 (Fig. 1). The genuine common-driving motif (illustrated in Fig. 2) is designated M3. Node 2 is the common driver whereas nodes 1 and 3 are the common-driven ones. In particular, we pay special attention to the cross-correlation between nodes 1 and 3. With the exception of illustrative time traces and their corresponding analysis, the results represent an average over 40 independent runs, unless otherwise stated, with different random initial conditions with error bars given by the corresponding standard deviation. We characterize the synchronization in other motifs that represent structural variations of the M3 motif: the addition of one or more connections (M6, M8, M9, M13), or the addition of connections and nodes (e.g., M3+1). In particular, M9 is the prototypal dynamical-relaying motif [24], which has been previously shown to promote zero-lag synchronization in a variety of systems [33-38], including neuronal systems [25-27].

**Common-driving motifs without and with resonance pairs**

We first focus on the four motifs depicted in Fig. 2. The simple common driving motif (M3), in which node 2 drives the dynamics of nodes 1 and 3 was contrasted with three other motifs (M6, M9 and M3+1), which represent structural variations of M3. Because motif M3 lacks any feedback or cyclical structure, the conduction delay plays no role in the dynamics or in the synchronization between nodes 1 and 3: Hence the outer nodes passively receive the driver's input. Onto this "backbone", motif M6 has a single feedback connection added, forming a reciprocal connection between nodes 1 and 2. Motif M9 has reciprocal connections between node 2 and nodes 1 and 3. Motif M3+1 possesses an extra node (4) reciprocally connected with node 2.

**Motifs of Hodgkin-Huxley neurons.** For the smallest-scale system we consider, each node comprises a single excitatory Hodgkin-Huxley neuron [39] weakly inter-connected with a conduction delay of 6 ms. Each neuron receives independent Poisson trains of spikes, representing background stochastic input. Stimulated by such external input, neurons exhibit continuous spiking behavior with average inter-spike interval of approximately 15 ms and are hence suprathreshold, regardless of the input from the other motif neurons. When the neurons are coupled according to the M3 motif, as shown at the top row of Fig. 2, spikes from the center neuron 2 only sporadically trigger simultaneous spikes of neurons 1 and 3 (following the common 6 ms delay). Panels a and b illustrate an exemplar time trace of the neurons. As a consequence of the absence of regular coincident spikes in the outer neurons, the maximum of the cross-correlation between nodes 1 and 3 is small, evident in both the single trial (Fig.

2 c), and average (Fig. 2 d) results. Note in particular that the cross-correlation function between the central neuron and an outer neuron has a modest peak corresponding to the 6 ms time delay (blue trace in third and fourth columns). The smaller peak at zero lag (red trace) reflects this common time delayed peak from the center to each of the two outer nodes. On the other hand, when the neurons are coupled according to the structural variations of the M3 motif (namely M6, M9 and M3+1), as shown in the second to the fifth rows of Fig. 2, spikes from neuron 2 reliably trigger simultaneous spikes in neurons 1 and 3. This is evident in the exemplar time series as well as the single and average cross-correlation functions. This is quite a striking change, given that all other parameters of the model remain unchanged from M3.

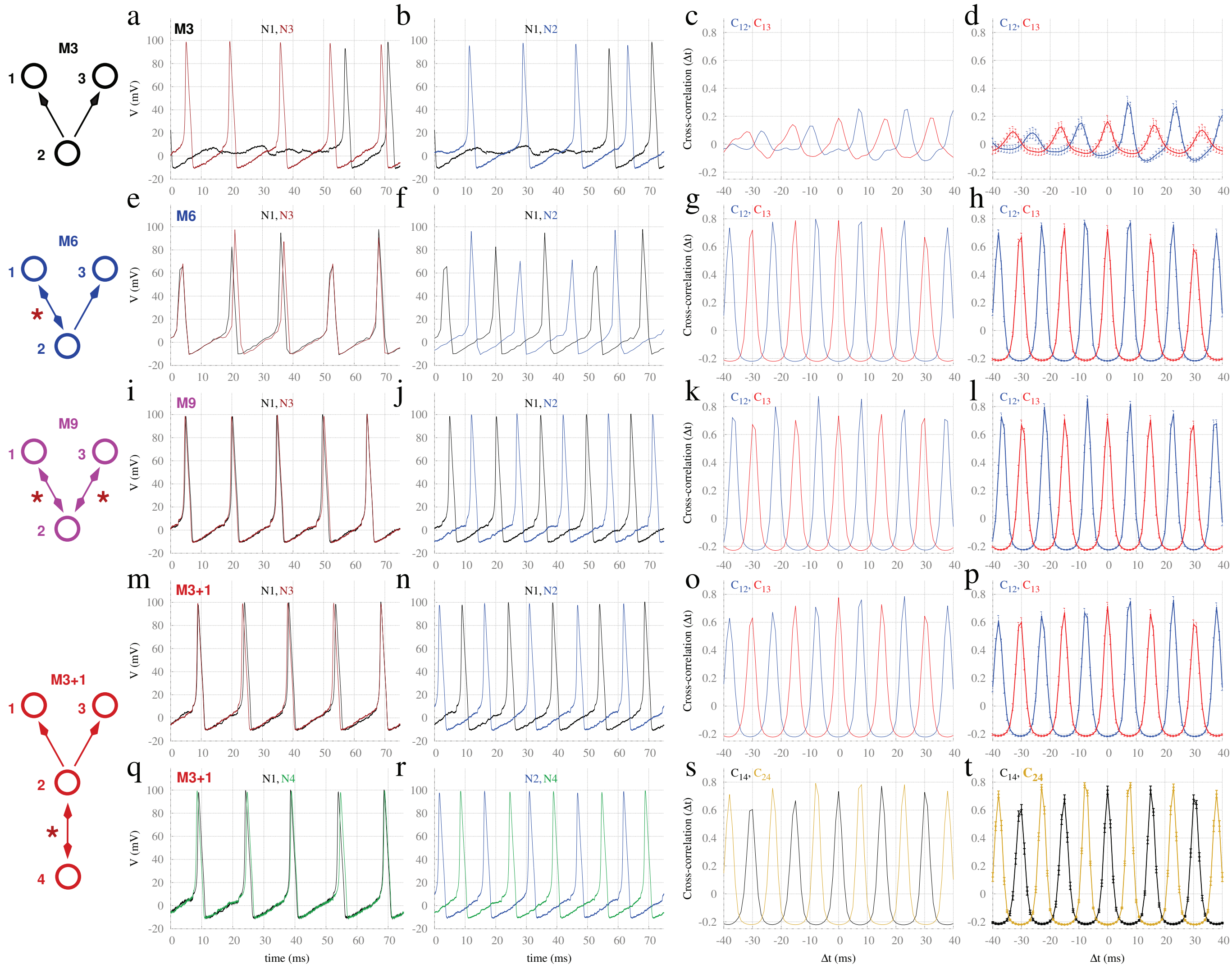

*Figure 2: Synchronization in motifs of Hodgkin-Huxley neurons. Dynamics of common driving motif (M3) versus common driving motifs with resonant sources (M6, M9 and M3+1) in motifs of excitatory delayed-coupled Hodgkin-Huxley neurons with delay $\tau$ = 6 ms. First and second columns (panels a, b, e, f, i, j, m, n, q, r) correspond to individual spiking time traces of neurons, whereas the third and forth columns (panels c, d, g, h, k, l, o, p, s, t) correspond to the cross-correlation functions of the corresponding single time series and average over 40 trials respectively. Descending rows show motifs M3, M6, M9 and M3+1, respectively.*

The structural variations introduced in motifs M6, M9 and M3+1 over the common driving M3 share an essential feature: The driver node 2 is mutually connected and synchronized with at least one other node. We denote this mutual connection *resonance pair*: Its presence dramatically alters the dynamics and synchronization properties of the driven nodes. Supplementary Fig. S1 compares the dynamics of two oscillators with different types of time-delayed coupling. Synchronization between these pairs appears exclusively when they are mutually coupled (in this particular case the synchronization is in anti-phase at the slow rhythm). Therefore, the resonance pair, identified by the red stars in the motifs, is the source of resonance-induced synchronization, leading to zero-lag synchronization between the outer nodes.

The emergence of zero-lag synchrony in motifs of coupled Hodgkin-Huxley neurons shows a strong dependence on the time delay, consistent with prior work [40]. This delay effect is crucial to the dynamics of motifs containing a reciprocal coupling, but not for the common driving M3. To compare the dynamics of motifs M3 and M6 it is instructive to analyze both the cross-correlations between pairs of nodes and the regularity of the inter-spike intervals (ISIs). To measure the irregularity of the inter-spike intervals, we use the incoherence R, defined as the coefficient of variation (CV) of the ISI, $R = STD(ISI)/\langle ISI \rangle$ [41,42], where STD stands for the standard deviation. Large values of R indicate more irregular patterns of ISIs. As shown in Fig. 3 (panel a), the incoherence of each node in M3 is independent of the delay and is larger for the driver node. In contrast, the incoherence of each node is similar in M6 and shows a strong effect on the time delay (top panels of Figs. 3 b-d), increasing and decreasing with a period of approximately half of the average ISI.

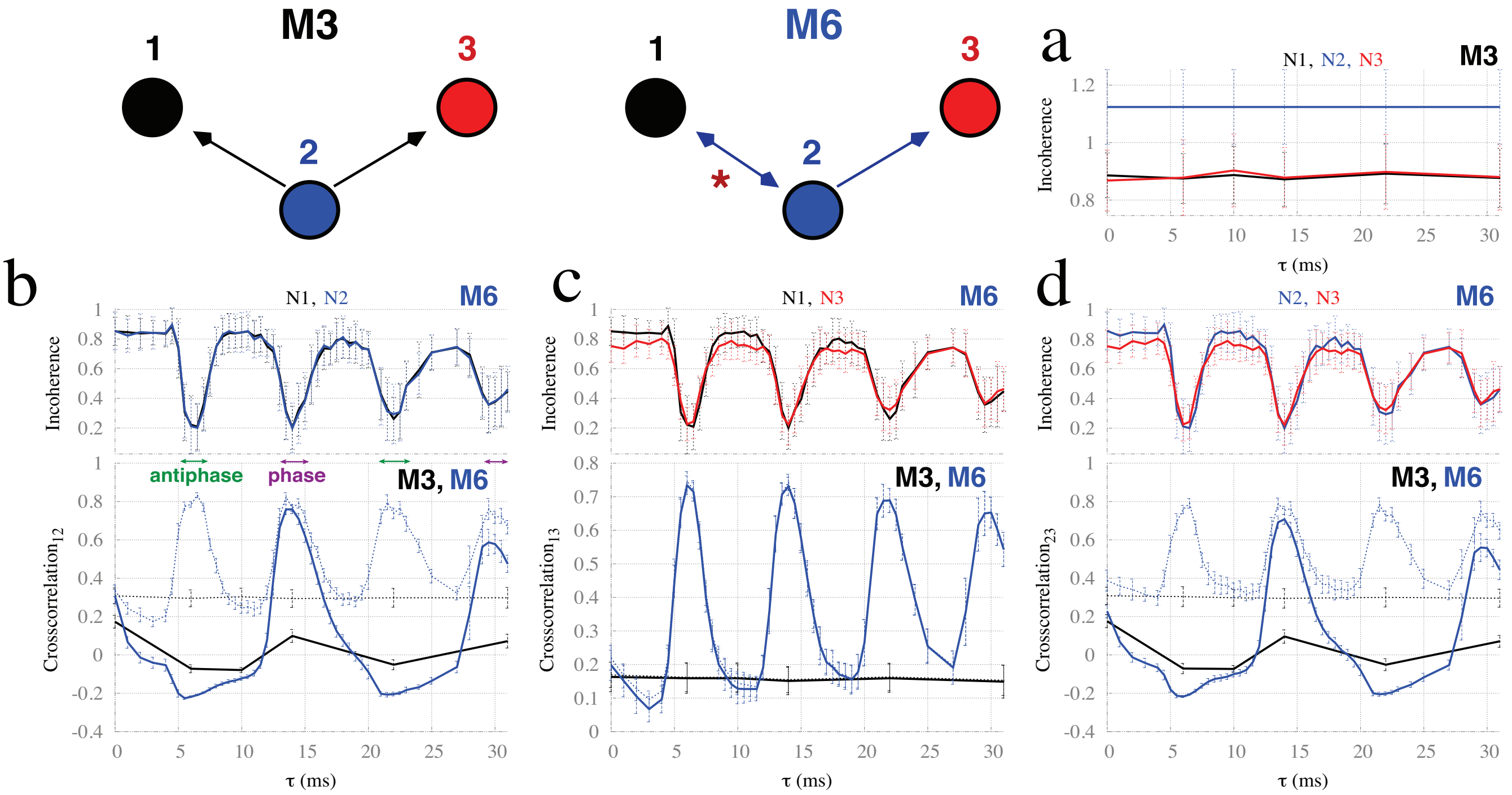

*Figure 3: Synchronization dynamics and incoherence in Hodgkin-Huxley neurons. (a) Incoherence in motif M3 does not depend on time delay. Colors indicate the different nodes. (b-d) Top panels show incoherence for M6, where colors represent different nodes, and bottom panels show crosscorrelations for M6 (blue) and M3 (black). Continuous lines indicate the cross-correlation coefficients at zero time lag, and dashed lines indicate the maximum cross-correlation coefficients for all time lags. Panels b, c and d represent pairs of nodes: 1-2, 1-3, and 2-3 respectively. Phase, anti-phase synchrony, and asynchrony can be found in motif M6 depending on the time delay τ (see exemplar time traces in supplementary Fig. S2). Results are averaged over 40 trials.*

The bottom panels of Figs. 3 b-d compare the cross-correlation between pair of nodes at zero-lag (continuous lines) against the maximum across all time lags (dashed lines) for motifs M3 (black) and M6 (blue). For motif M3, the maximum cross-correlation does not depend on the time delay. The input from node 2 solely arrives at nodes 1 and 3 after different latency times, but this delay does not impact on the dynamics of the driven nodes. In contrast, for motif M6 the maximum cross-correlations (blue dashed lines) vary, with peaks that coincide with the minima of incoherence (top panels). The synchronization in the Hodgkin-Huxley model, which has only one oscillatory frequency, appears either in phase or in antiphase. For neighboring nodes (1-2 or 2-3), phase synchronization occurs when a peak of the maximum cross-correlation coincides with a peak of the cross-correlation at zero lag. Anti-phase synchronization occurs when a peak of the maximum cross-correlation coincides with a minimum of the cross-correlation at zero lag. Supplementary Fig. S2 illustrates example time traces of Hodgkin-Huxley neurons in M6 for anti-phase synchronization (τ=6 ms), no synchronization (τ=10 ms), and phase synchronization (τ=14 ms).

The delay in a resonance pair can either enhance or reduce the synchronization. In motif M6 the driven nodes (1 and 3) synchronize whenever the resonance pair (1 and 2) synchronizes, whether this is in-phase or anti-phase synchrony. This corresponds to a drop in the incoherence R of the driving node. Thus, the synchronization between nodes 1 and 3 depends on the time delay in the resonance pairs because the synchronization between the resonance pair (1 and 2) - and thus the incoherence - also depends on the time delay. It thus appears that synchronization between the reciprocally connected nodes leads to a more regular (less incoherent) output from the master node which then facilitates synchronization between this node and the other slave node.

**Motifs of neuronal populations.** To investigate whether these results translate to neuronal activity at the next spatial scale, we exploited the computational parsimony of the neural model of Izhikevich [43, 44] to study populations of spiking neurons, each node comprised a population of (400) excitatory and (100) inhibitory randomly interconnected neurons [27], with each neuron in these populations receiving an independent Poisson spike train. Neurons of the same populations were synaptically coupled without conduction delay and with a latency of 15 ms for (exclusively) excitatory inter-population connections. We focused on the dynamics of the ensemble mean membrane potential $\langle V \rangle$ of all neurons within each population. As shown in the time traces of Fig. 4, the activity of each population consists of two time scales, a higher frequency (≈ 25 Hz) brief network spikes and a lower frequency fluctuation (≈ 3 Hz) on which these transients typically recur. Notably, the dominant (low frequency) time scale – which does not occur in the single neuron system - is much longer than the conduction delays. Despite discrepancies in the time scales and nature of the dynamics, the zero-lag synchronization reported in Fig. 4 largely resembles that shown in Fig. 2. Dynamics on the common driving motif M3 between the central and outer nodes show a moderate time delayed cross-correlation ($C_{12}{\sim}0.4$) at approximately 20 ms and a corresponding weak to moderate zero-lag synchrony between the outer nodes ($C_{13}{\sim}0.3$). However zero-lag synchrony is substantially stronger on the motifs possessing at least one resonance pair (M6, M9, M3+1). Notably, the anti-phase relation between node 2 with respect to nodes 1 and 3 appears solely at the faster time scale, comparable to the delay period.

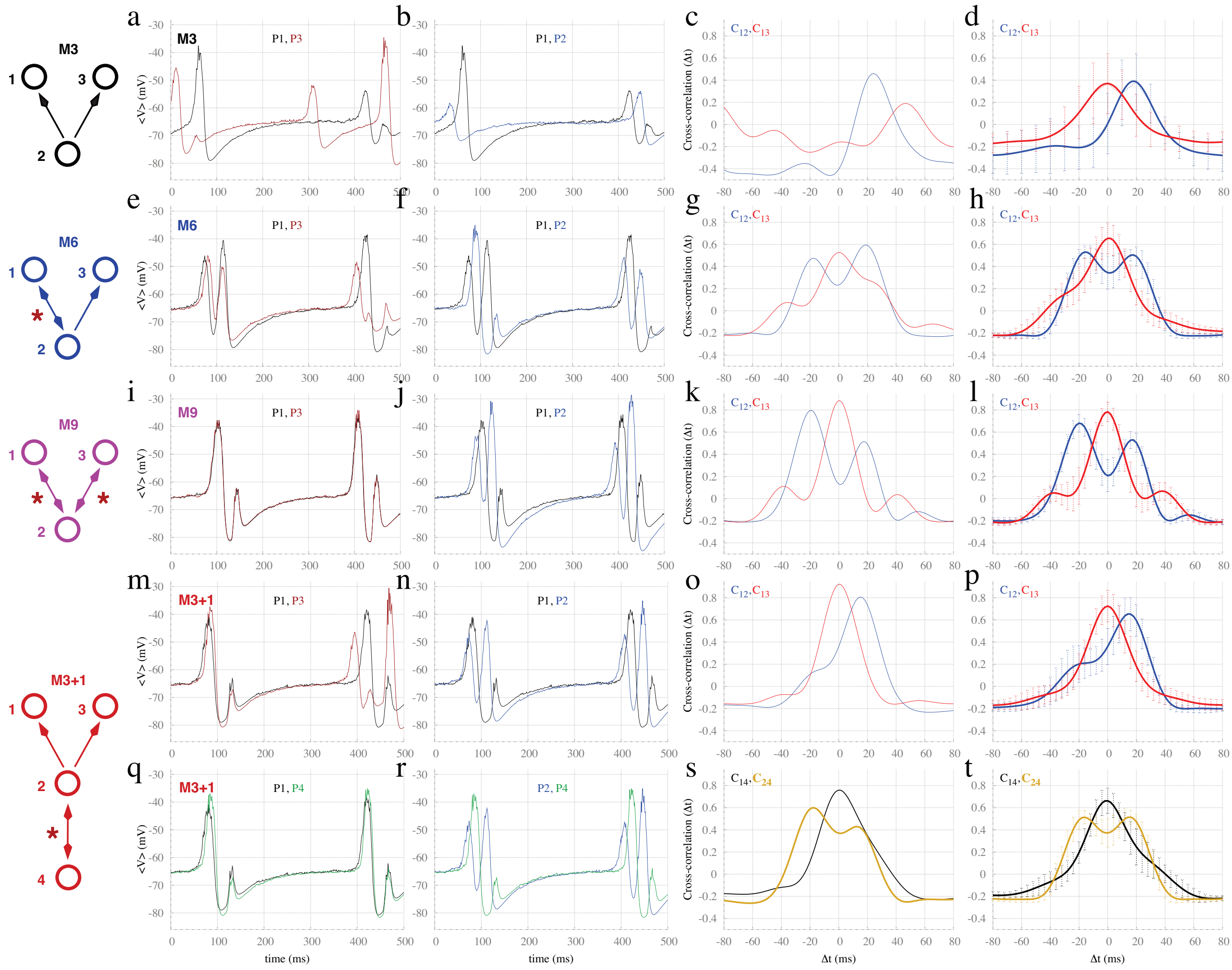

*Figure 4: Synchronization in motifs of populations of Izhikevich neurons. Panels (a-t) as per Fig. 2, for populations of 500 (400 excitatory and 100 inhibitory) spiking neurons and delay τ = 15 ms.*

Similar to motifs of Hodgkin-Huxley neurons, synchronization of populations of spiking neurons also shows a dependence on the delay time between nodes in the presence of a resonance pair. Supplementary Fig. S3 shows that the incoherence (here using the inter-burst interval instead of the inter-spike interval) and cross-correlations in motif M3 do not depend on the delay (panel a). However, they vary considerably for motif M6. Supplementary Figs. S3 b-d show that large incoherence values for motif M6 correspond to the transition between the regimes of phase and anti-phase synchronization for neural populations. Supplementary Fig. S4 illustrates two cases of synchronization with one dominant oscillatory frequency (one in phase with $\tau$=8 ms, and another in anti-phase with $\tau$=32 ms), and one case ($\tau$=20 ms) of synchronization with two oscillatory frequencies that is in phase for the slow rhythm and in anti-phase for the fast rhythm. For any time delay, the synchronization between 1 and 3 is enhanced for motif M6 when compared to M3, and the synchronization between nodes 1 and 3 largely resembles the maximum synchronization between first neighbors (nodes 1 and 2, or 2 and 3).

**Motifs of neural mass models.** To further study the robustness of the relationship between motifs and synchronization with respect to the underlying dynamical systems, we next utilized a neural mass model, which represents a reduced model of cortical dynamics. A neural mass model is a parsimonious representation of the dynamics of a very high-dimensional system, and replaces thousands of equations for each population of neurons with a small number (here only three) of nonlinear equations per node. These represent the dynamical behavior of the essential summary system statistics (here mean firing rate) and hence a reduced representation of spontaneous cortical dynamics. Here we employ a population representation of conductance-based model neurons [5,6,45,46], as has been previously used to elucidate important features of large-scale brain dynamics [47,48]. This system also breaks from the previous two scales studied above in that irregularity is dynamically generated (through endogenous chaotic dynamics within each mode) rather than introduced through external stochastic spikes.

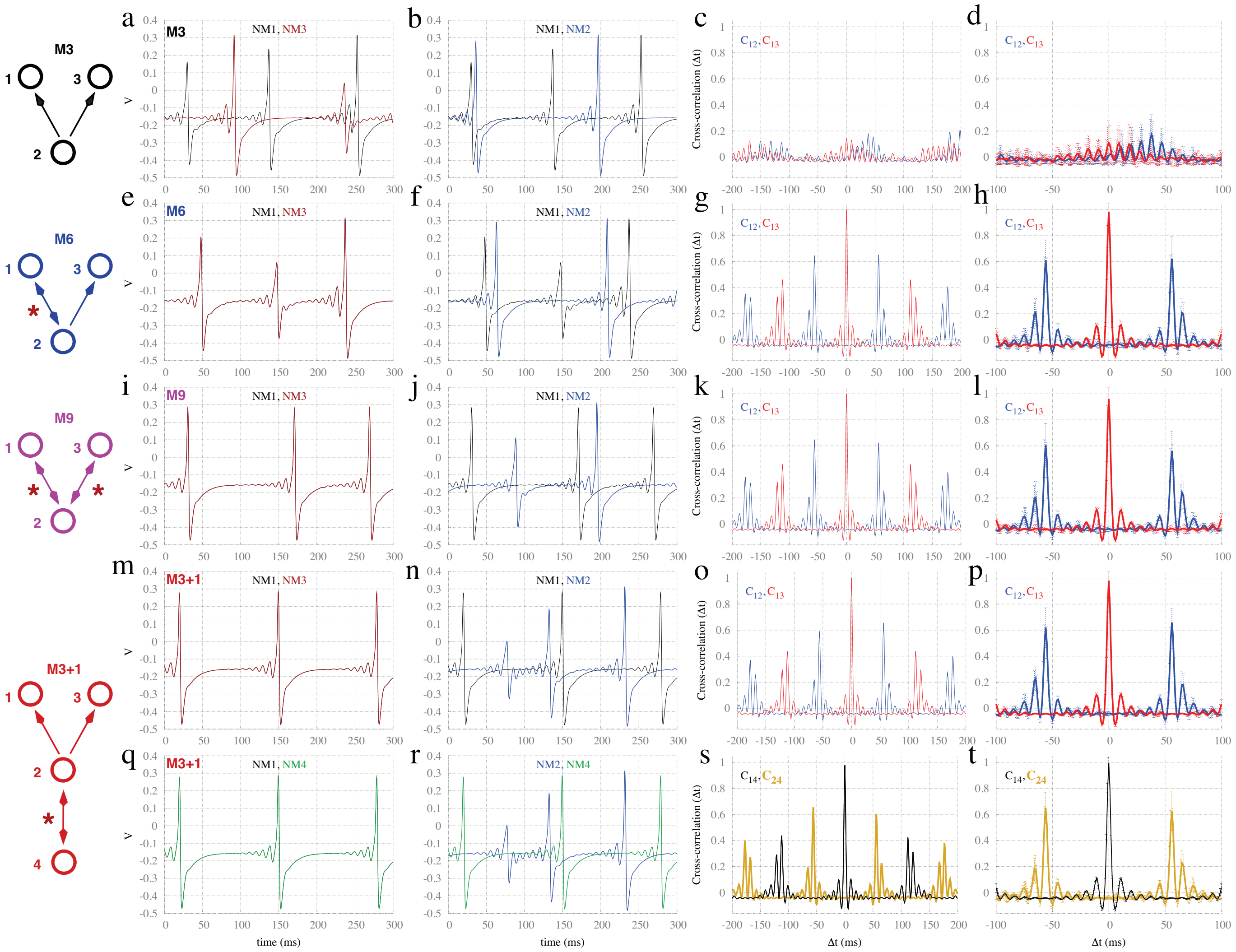

*Figure 5: Synchronization in motifs of neural mass models. Panels (a-t) as per Fig. 2, but for neural mass models with coupling strength c = 0.01, and delay τ = 10 ms.*

The dynamics of these delayed-coupled neural masses shows chaotic oscillations fast dynamics (~100 Hz) superimposed on slower return times of about 110 ms [6]. As shown in Fig. 5, the dynamics in this system clearly replicate those observed above, namely that zero-lag synchrony between nodes 1 and 3 was strongly and exclusively expressed in the motifs with resonant sources (M6, M9, M3+1). It can be seen that within the resonance pairs, node 2 is in anti-phase synchrony with nodes 1 and 3. Notably, however, the anti-phase relation typically occurs at a much slower time scale (110 ms) than the coupling delay (10 ms).

Despite the dissimilarities between the neuronal systems at different scales, synchronization and incoherence of the neural mass model also exhibits a dependence on the time delay in the presence of a resonance pair (see supplementary Fig. S5). To better understand synchronization dynamics in this system - which has multiple internal time scales - it is necessary to study the combination of time delays *and* coupling strength: For weak coupling strength (c=0.01), phase synchronization is not reached. However, as illustrated in supplementary Fig. S6, rich dynamics can arise, including anti-phase synchronization at the slow ($\tau$=0 ms) or fast ($\tau$=75 ms) time scales, or an asynchronous state ($\tau$=35 ms). In contract, when the coupling is stronger (Supplementary Figs. S7, S8, left), phase synchronization emerges for very short time delays. In the cases of stronger coupling, for example c=0.05 or c=0.15 (Figs. S7, S8, right), zero-lag synchronization between nodes 1 and 3 is also more stable for long delays.

**Mismatch in the conduction delays**

Biological systems are naturally diverse, and therefore, any relevant behavior should not be highly dependent on the fine-tuning of the delay – and particularly its symmetry. We next tested the generality of the zero-lag synchronization between nodes 1 and 3 with respect to delay mismatch in the different motifs containing the resonance pair. The connections preserved the conduction delay of $\tau$ except for a single feedback connection to the driver node 2 in motifs M6', M9' and M3+1' in which we introduced a variable conduction delay in one direction ($\tau$'), as illustrated in Fig. 6 a. The three motifs exhibited zero-lag synchronization that was substantially larger than that of motifs M3 (black line) or even M3 plus a unidirectional input (yellow line) across a large region of the parameter space (Fig. 6 b-d). In the motifs of Hodgkin-Huxley neurons (Fig. 6 b), the behaviors of all three motifs are similar for $\tau' < \tau$. In contrast, for $\tau' > \tau$ zero-lag synchrony decays in a similar way for motifs M6' and M3+1', whereas synchronization in motif M9' is virtually independent of $\tau$' for up to fivefold $\tau$ (not shown in the plot). Supplementary Fig. S9 shows the analyses of the dynamics of motif M6' in more detail: It shows that synchronization arises in M6' only when the delay mismatch $\tau$' yields synchronization with the same phase relation as $\tau$, which - in the case of $\tau$=6 ms - is anti-phase synchronization between neighboring neurons (see Fig. 3). The motifs of neural mass models show a systematic consistency of synchronization across $\tau$' for a biologically plausible range of delays (Fig. 6 c). However, a behavior similar to that observed in motifs of Hodgkin-Huxley neurons occurs for greater delay mismatches (Fig. 6 d). Such differences in the time scales are consistent with the different time scales of these systems: The Hodgkin-Huxley neurons oscillate with periods of about 15 ms, whereas the neural masses oscillate with periods of about 110 ms.

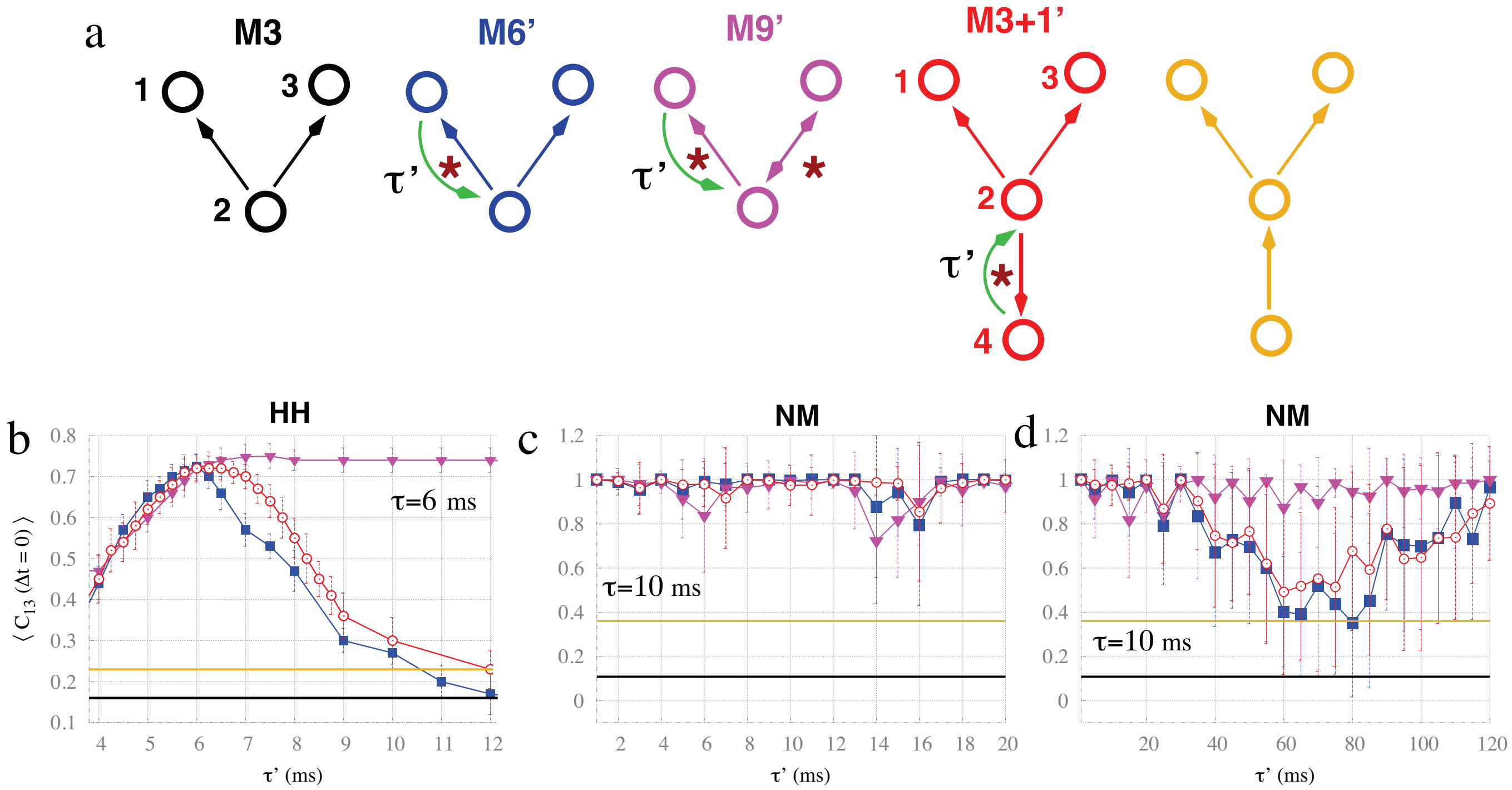

*Figure 6: Robustness of the synchronization with respect to mismatch in the delays. The top schemes (a) illustrate the motifs of neurons considered. Motifs M6', M9' and M3+1' have one connection with delay τ', and all the other connections have delays of 6 ms. The bottom panels show the zero-lag cross-correlation between nodes 1 and 3 in motifs of Hodgkin-Huxley neurons (b) and in motifs of neural mass models with c=0.01 (c) averaged over 40 trials for varying τ'. Panel (d) shows the same as (c) but across a broader range of τ'. Plot colors correspond to motifs as per panel a.*

## Characterizing the dynamics of the motifs

From herein, we focus on motifs of neural masses, exploiting their relative computational parsimony to gain deeper insight into the mechanisms of the resonance pair. In particular we studied the robustness of our findings with respect to the most salient parameters of the system, namely the coupling strength and the delay. As shown in Fig. 7, the strength of the synchronization in the motifs with a resonance pair, but not M3, show an increase as a function of coupling strength (panels a, b). Although an expected feature of the model [49], the emergence of synchrony even at very weak coupling ($c \sim 10^{-3}$) is somewhat surprising for a biological system. There are, however, some regions of complex dynamics (evidenced as large error bars) in which there is not a unique solution, thereby entailing significant trial-to-trial variability. At relatively weak coupling (c=0.01), zero-lag synchronization between nodes 1 and 3 holds across a broad regime of physiologically plausible time delays (Fig. 7 c). Analysis of longer coupling delays (supplementary Fig. S10) reveals an influence on synchronization that resembles the system of Hodgkin-Huxley neurons (Fig. 3), albeit weaker and at a much longer time scale.

These analyses suggest a partition of the common-driving motifs into three distinct families: (i) The simple common-driving motif (M3) where synchronization at zero lag is not achieved in the weak-coupling regime, independent of the time delay; (ii) A ring of three mutually coupled systems (M13) or a common-driving motif that also contain direct coupling between the driven nodes (M8) require a relative strong coupling and negligible delay in order to promote synchronization (Fig. 7 d-f), because of the existence of frustration; and (iii) Common-driving motifs enhanced by active resonance pairs (e.g., M6, M9, M3+1) which exhibit zero-lag synchronization even for very small couplings, irrespective of the time delay (up to $\tau$=20 ms). It is clear in these analyses that the increase in zero-lag synchrony in motifs with a resonance pair is not due to the additional coupling introduced by the backward connection, but rather through the placement of the additional edge. For example, the motifs with the greatest number of edges (M8 and M13) are amongst the most difficult to achieve zero-lag synchrony with an increase in coupling. Closing the outer nodes with two additional edges (going from M9 to M13) leads to a substantial decrease in zero-lag synchrony.

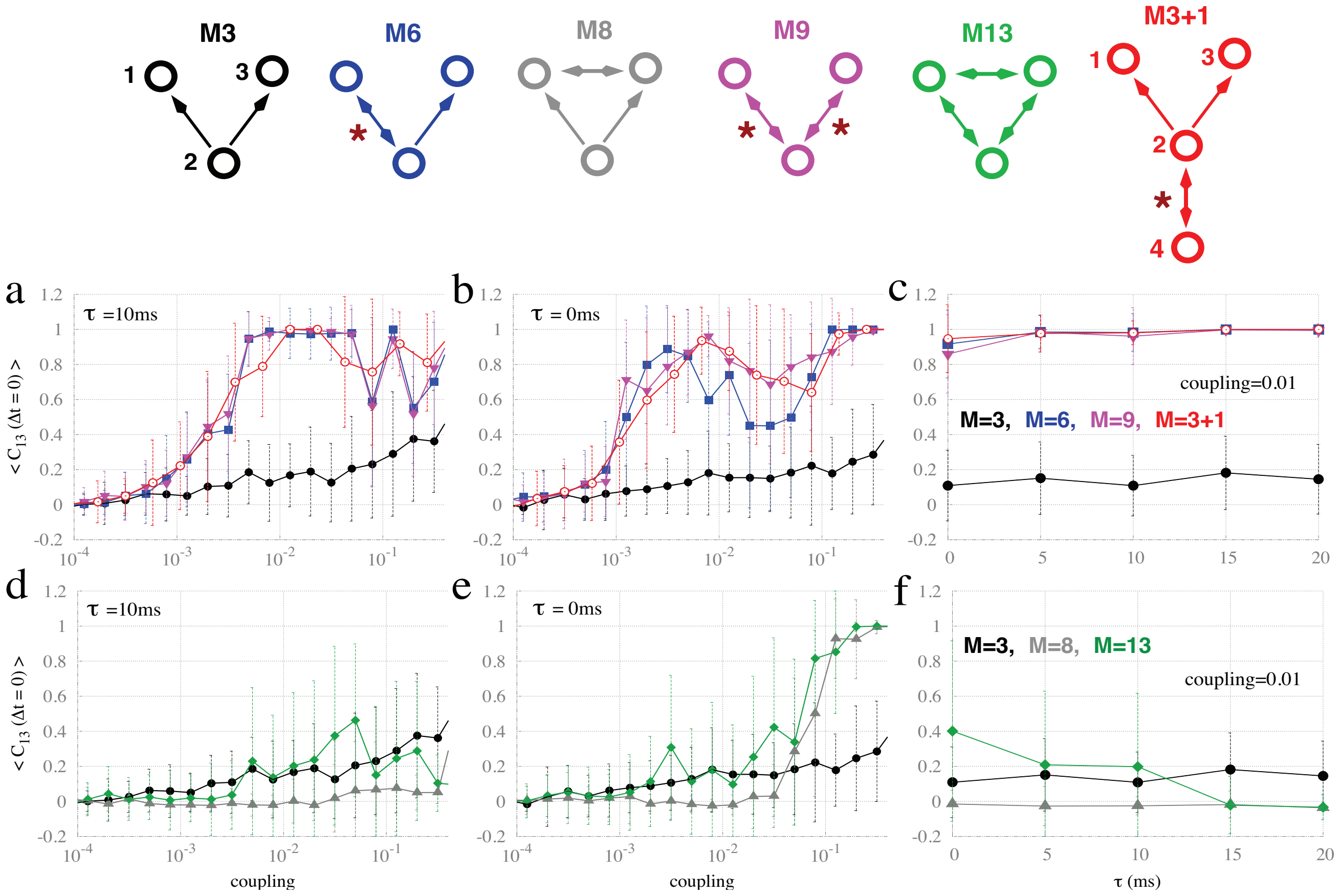

*Figure 7: Common-driving motifs, labeled as per Sporns and Kötter (2004) [9], see Fig. 1 a. Bidirectional connections (red stars) indicate active resonance pairs. Zero-lag cross-correlation between neural masses 1 and 3 for the six motifs studied. Top row compares common driving (M3) to common driving with resonance pairs (M6, M9 and M3+1) for varying coupling (panels a and b) and varying delay (panel c). Bottom row compares common driving (M3) to common driving with a distributed resonance pair (M13), and to common driving plus a bidirectional connection between 1 and 3 (M8) as a function of coupling (panels d and e) and time delay (panel f).*

**Propagation of the effect of the resonance pair**

The preceding analyses show that the effect of the resonance pair can influence the common driving motif even when it is placed outside the motif itself (e.g. M3+1). Here we further investigate the propagation of the resonance pair effect by considering larger structures in which the resonance pair is distant from the driver node (2). This procedure is schematically shown in Fig. 8 a, and illustrated for a particular network of N=7 nodes in Fig. 8 b. We are particularly interested to understand if the effects of the resonance pair are strictly local, and, additionally, on how the polysynaptic distance to the resonance pair influences the dynamics and synchronization.

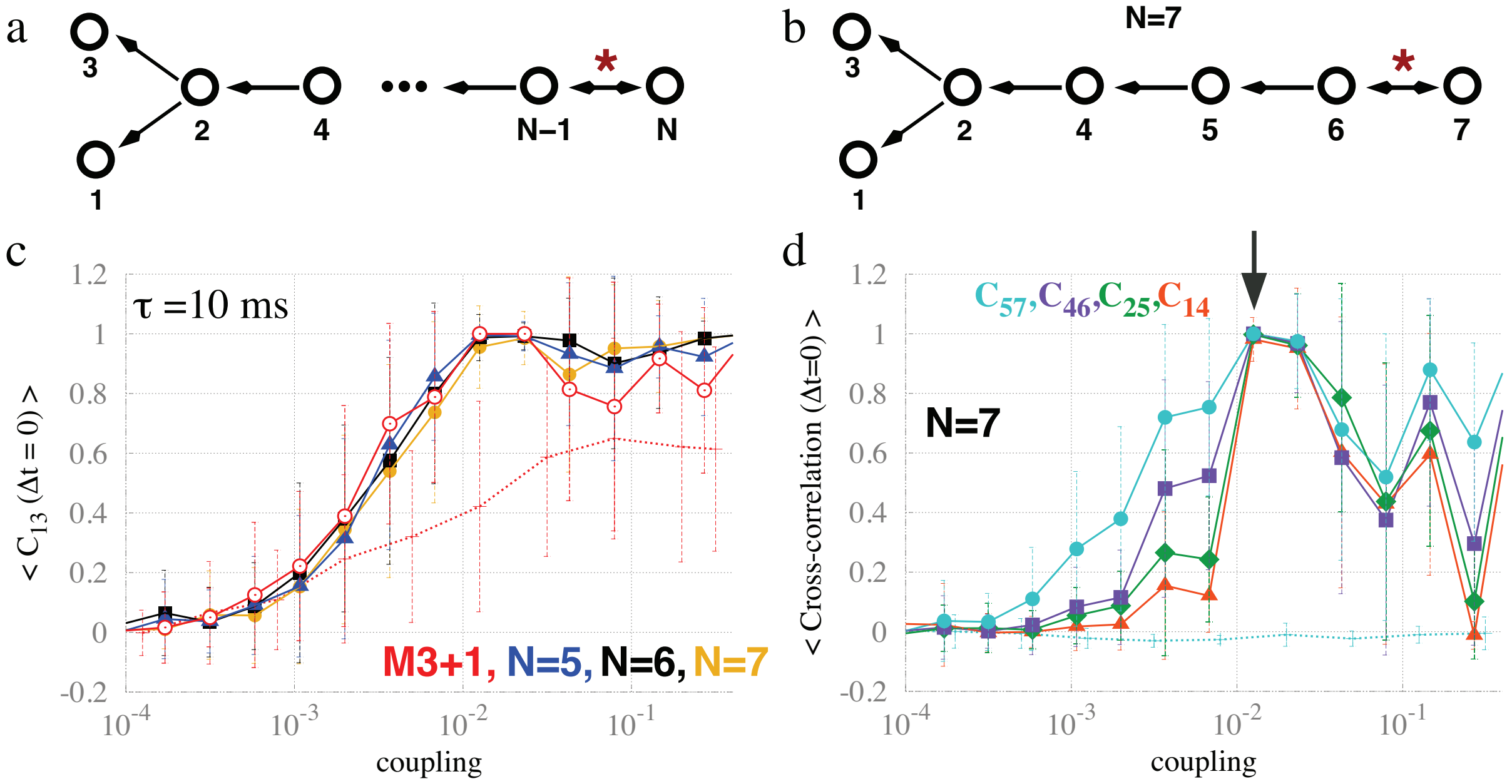

*Figure 8: Propagation of the effect of a resonance pair along a chain. (a) A resonance pair (nodes N and N-1) arbitrarily distant from a pair of commonly driven neural masses (1 and 3). (b) A chain with 7 nodes. (c) Zero-lag cross-correlation functions between nodes 1 and 3 for different chain sizes as illustrated in panel (a) are shown in solid lines. Thin dashed line represents the chain of panel (a) without the feedback connection from node N-1 to node N. (d) Zero-lag cross-correlation functions between every other node in the chain depicted in panel (b) are shown in solid lines. Thin dashed line represents the chain of panel (b) without the feedback connection from node 6 to node 7.*

We observe that zero-lag synchronization between the driven nodes 1 and 3 is virtually independent of the distance along a polysynaptic chain from the resonance pair (Fig. 8 c). For a fixed motif length (N=7), we also characterized the zero-lag synchronization of different pairs of nodes that did not interact directly, but interacted indirectly through a common neighbouring mediator (see Fig. 8 d). Apart from pairs 5-7, all such pairs correspond to a strict flux of information flow, mandated by the direction of the coupling. Thereby, the synchronization decreased with the distance from node 7, unless the system was set with a specific coupling (see arrow in Fig. 8 d) that gives rise to global synchronization. This corresponds to identical synchronization between nodes 2, 5 and 7, which are anti-phase synchronized to nodes 1, 3, 4 and 6 occurring at this particular coupling strength.

Finally, to highlight the influence of the resonance pair in the dynamics, we removed the feedback connection to node N (results shown as thin dotted lines in Figs. 8 c and d). By means of this control simulation, we find that: (i) Zero-lag synchronization between 1-3 is consistently reduced (Fig. 8 c); and (ii) Zero-lag synchronization between 5-7 (Fig. 8 d) completely disappears in the absence of a resonance pair.

### Characterizing active resonance pairs

We have denoted an active resonance pair as two mutually connected nodes that synchronize in the presence of appropriate time delays and coupling strength. This effect propagates through the motifs because the driven nodes show a strong tendency to synchronize with the driver node (hence promoting zero-lag synchronization between driven nodes). That is, the emergence of synchronization between the resonance pair then stabilizes synchrony amongst undirectionally coupled nodes. The same phenomenon underlies the propagation down a polysynaptic chain (Fig. 8). Interestingly, the impact of the resonance pair extends beyond this propagation, giving rise to other dynamical effects for coupling delays in which anti-phase synchrony between neighbors prevails. Geometrical frustration is an example: In some motif configurations, anti-phase synchrony between pairs of mutually connected nodes (potential resonance pairs) is simply not a stable solution. In the case of motif M13 (illustrated in Fig. 7), for example, anti-phase synchronization between any pair is frustrated because the third node cannot be simultaneously synchronized in anti-phase with respect to the other two neighbor nodes. This situation illustrates that frustration can disturb potential resonance pairs. Large mismatches in the delays of the mutual connection between the pair can also disturb the effects of a resonance pair. As depicted in Fig. 6, both motifs M6' and M3+1' are similarly susceptible to mismatches in the reciprocal latencies.

### Transient behavior and the stability of synchronization in resonance-pair motifs

Connectivity also plays a role on the onset of synchronization. We studied the temporal onset of zero-lag synchronization in neural mass models for different motifs by (1) examining the transient dynamics following random initial conditions, and (2) studying the response to a transient perturbation. An example is shown in Figs. 9 a, in which dynamics on M6 begin from random initial conditions, then approach synchronization between masses 1 and 3. The dynamics are then perturbed by a brief current from 800 to 1000 ms - that is distinct for each driven node - before rapidly regaining synchrony after a few hundreds of milliseconds. It is noteworthy that the approach to zero-lag synchrony in both scenarios is approximately exponential, with an exponent $\gamma$ that can be used as a numerical estimate of the stability of the synchronous state (Figs. 9 b). In contrast, edge nodes on motif M3 remain unsynchronized. The dependence of the exponent γ with the coupling strength for the 1200 ms following offset of the transient perturbation is shown in Figs. 9 c. Motifs with resonance pairs (M6 and M9) showed a negative exponent, consistent with stable synchrony, whereas the exponent associated with motif M3 was positive throughout. Interesting, the coupling strength associated with the strongest synchrony (most negative exponent) occurred for a relatively weak coupling strength of *c*=0.01.

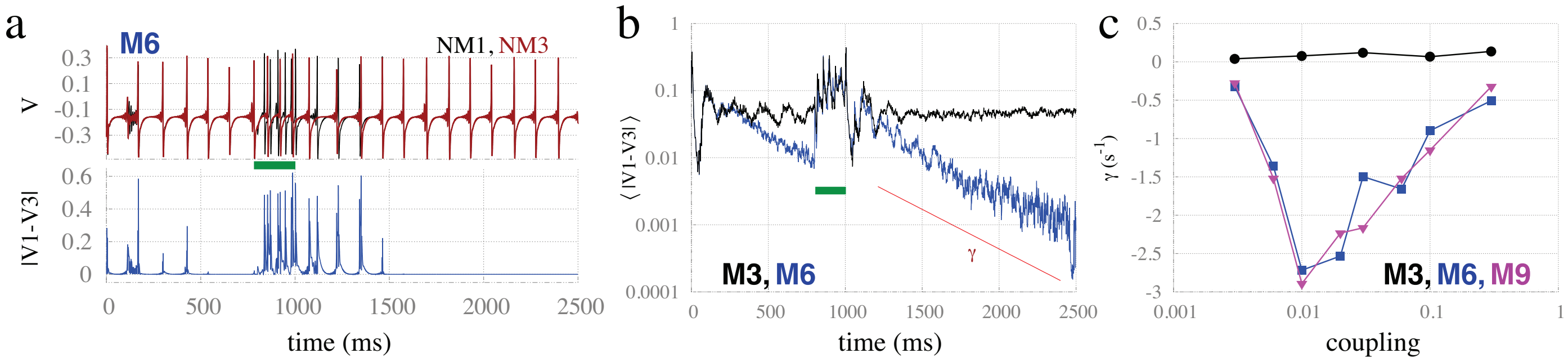

*Figure 9: Fast transient behavior and onset of synchronization. (a) Example of time-trace synchronization following random initial conditions (starting at time=0) and consequent to a brief perturbing current (green bar) at time=1000 ms in motif M6 with c=0.01. (b) |V1-V3| averaged over 400 trials with c=0.01 in motifs M3 compared to M6. (c) Exponent γ estimated from |V1-V3| averaged over 400 trials on the interval between 1200 and 2400 ms for varying coupling strengths in motifs M3, M6 and M9. Delay τ=10 ms.*

**Fine-tuning and synchrony in the absence of resonance pairs**

Synchronization hence arises quickly in the presence of a resonance pair. Is it possible to adjust the dynamics of the driver node without such reciprocal coupling to induce synchronization? We next studied this possibility by fine-tuning the input current ($I_0^2$) to the driver node (2) in motif M3, whilst keeping all other parameters fixed. As shown in Figs. 10 a, introducing a slight mismatch in the input current can indeed lead to large changes in the zero-lag synchronization between nodes 1 and 3. Crucially, careful fine-tuning of this current mismatch can lead to a near complete synchronization in motif M3 (A), or at least lead to a strong enhancement of synchronization (B and C). As depicted in Figs. 10 b, the maximum synchronization (A) occurs when the input current causes the driver node to exhibit the same oscillatory frequency as the driven edge nodes. The other local maxima occur when the driver node oscillates with a frequency that is an integer multiple of the driven nodes (2:1 in B and 3:1 in C). In contrast to this need for fine-tuning in motif M3, the resonance pair guarantees that node 2 oscillates with the same frequency as the driven nodes, with strong synchronization hence arising regardless of the coupling strength, as shown in Figs. 10 c for motif M3+1.

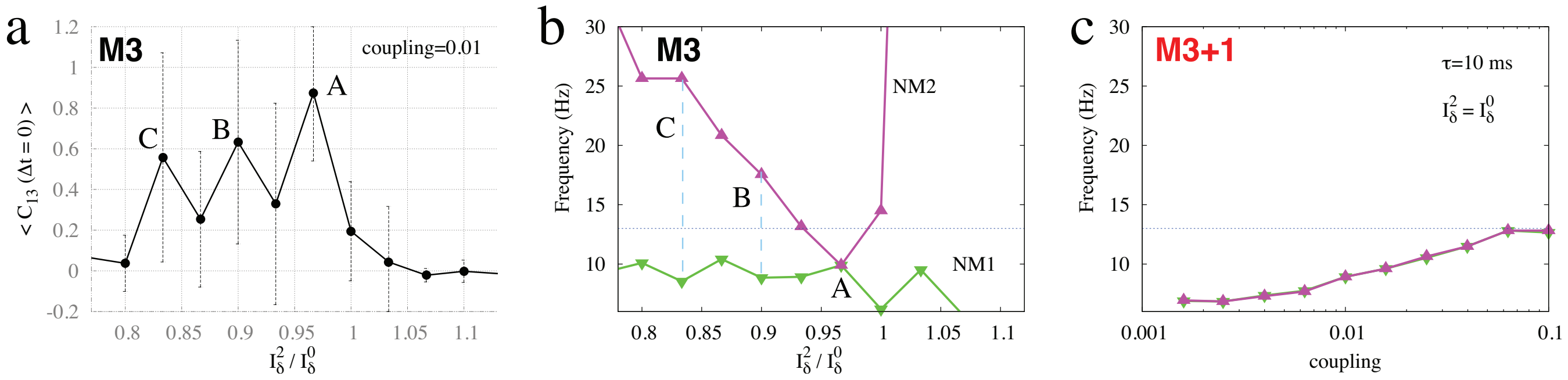

*Figure 10: Fine-tuning can enhance synchronization. (a) Crosscorrelation averaged over 40 trials, (b) dominant oscillatory frequencies of neural masses 1 (green) and 2 (magenta) as a function of the mismatch on the input current over node 2. (c) Dominant oscillatory frequencies of neural masses 1 (green) and 2 (magenta) for varying coupling strength.*

**Beyond resonance pairs**

The effects of a resonance pair can enhance the synchronization locally and even propagate in a polysynaptic way to influence distant dynamics. Reciprocally connected nodes can also interact in a way that disturbs the synchronization if they introduce frustration as in motifs M8 and M13, as shown in Fig. 7. To more deeply understand the role of reciprocally connected nodes and loops, we studied resonance motifs that go beyond the resonance pairs. Starting with a common driving motif M3, we added chains of bi- or uni-directionally coupled nodes of varying sizes as shown in Figs. 11 a. Adding one node reciprocally connected to node 2 recovers the resonance pair, which is clearly a more effective way of synchronizing the driven nodes than adding one extra unidirectionally connected node (the blue dashed line of Figs. 11 b). The addition of two reciprocally connected extra nodes in a closed loop (resonance triplet) had an effect that was analogous to the resonance pair, and again far more effective than the counterpart of two extra unidirectionally connected nodes in a loop (green dashed line). The addition of three or more reciprocally connected extra nodes in closed chain had a similar effect to the resonance pair. However, the influence of the unidirectionally coupled loops gradually approaches that of their reciprocally connected counterparts, which have already attained the ceiling effect (magenta dashed line). Hence, the interaction of unidirectionally connected nodes in a loop gradually enhances the synchronization of the driven nodes as the size of the loop increases. Therefore, even in the absence of reciprocally connected nodes, synchronization between 1 and 3 can be enhanced by a loop of at least three extra nodes connected to the driver node. Interestingly, the addition of a single resonance pair is the most efficient means of achieving zero-lag synchronization compared to loops of any size.

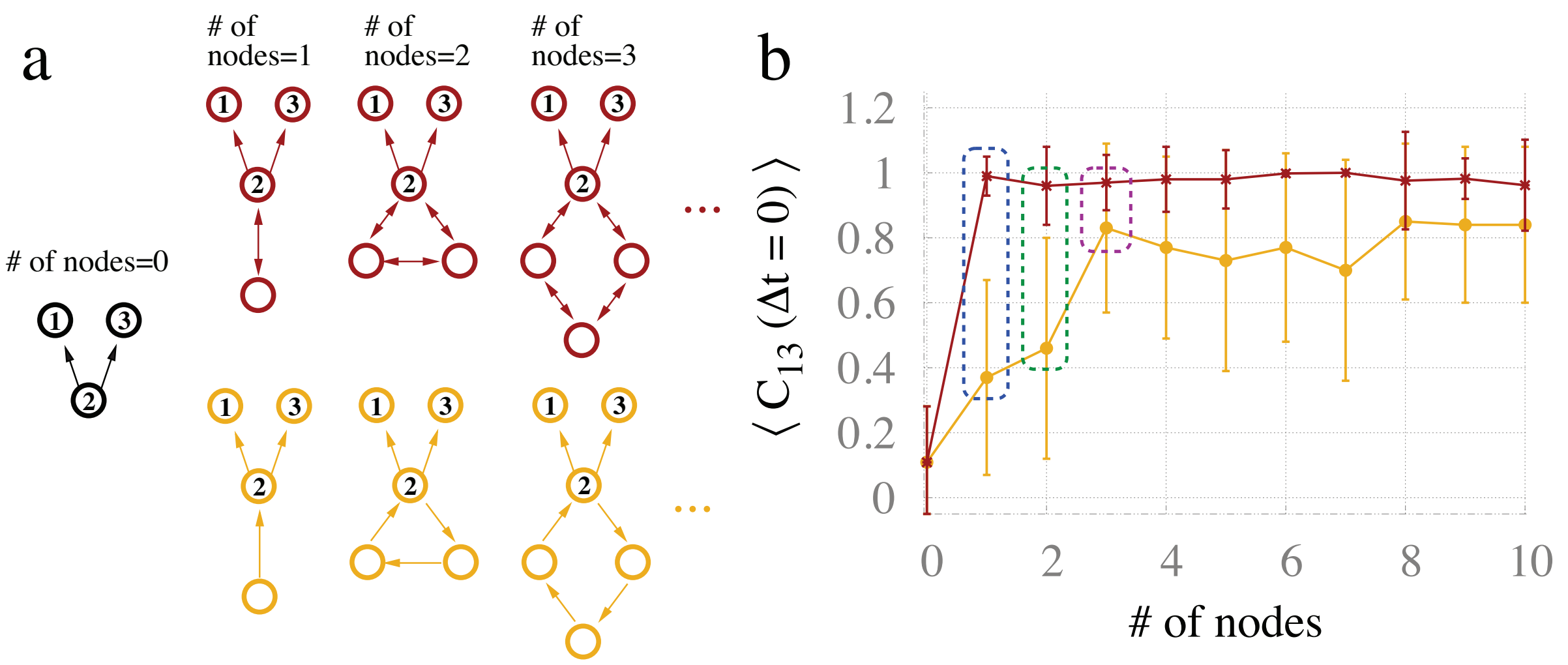

*Figure 11: Effect of resonance chains on the synchronization. (a) Loops of reciprocally connected versus unidirectional connected loops. (b) Zero-lag cross-correlation between neural masses 1 and 3 with neural mass 2 connected to bidirectional or unidirectional chains of varying length. Blue dashed line highlights the effect of the resonance pair, and green (magenta) dashed line highlights the effect of the resonance triplet (quad). Red (yellow) curve represents the cross-correlation averaged over 40 trials for reciprocally (unidirectionally) connected loops. The coupling strength is 0.01, and delay 10 ms.*

**Effects of the common driving input at higher orders**

Our final analysis concerns the synchronization properties of commonly driven nodes with higher polysynaptic orders (Figs. 12 a-c). In particular, we study the synchronization of the symmetrically located nodes n-n' for the different connectivity states of the driver node A. Figure 12 a illustrates the case in which node A was part of a resonance pair together with node B; Figure 12 b illustrates the case in which node A received a unidirectional input from node B; Figure 12 c illustrates the case in which node A did not receive input from any neighbouring regions. It can be seen in Figs. 12 d-g that only the motifs with the resonance pair (red line) yielded high correlation between nodes n and n' (for *n*=1,2,3,4). Interestingly, when the coupling strength is fixed (c=0.024) and the number of elements further increased (Fig. 12 h), the cross-correlation coefficient remained quite high for the chain containing the resonance pair. A similar behavior occurred for the maximum cross-correlation coefficient (for all time delays) between node A and node n (Fig. 12 i): Again, the resonance pair was required for the propagation of synchronous activity.

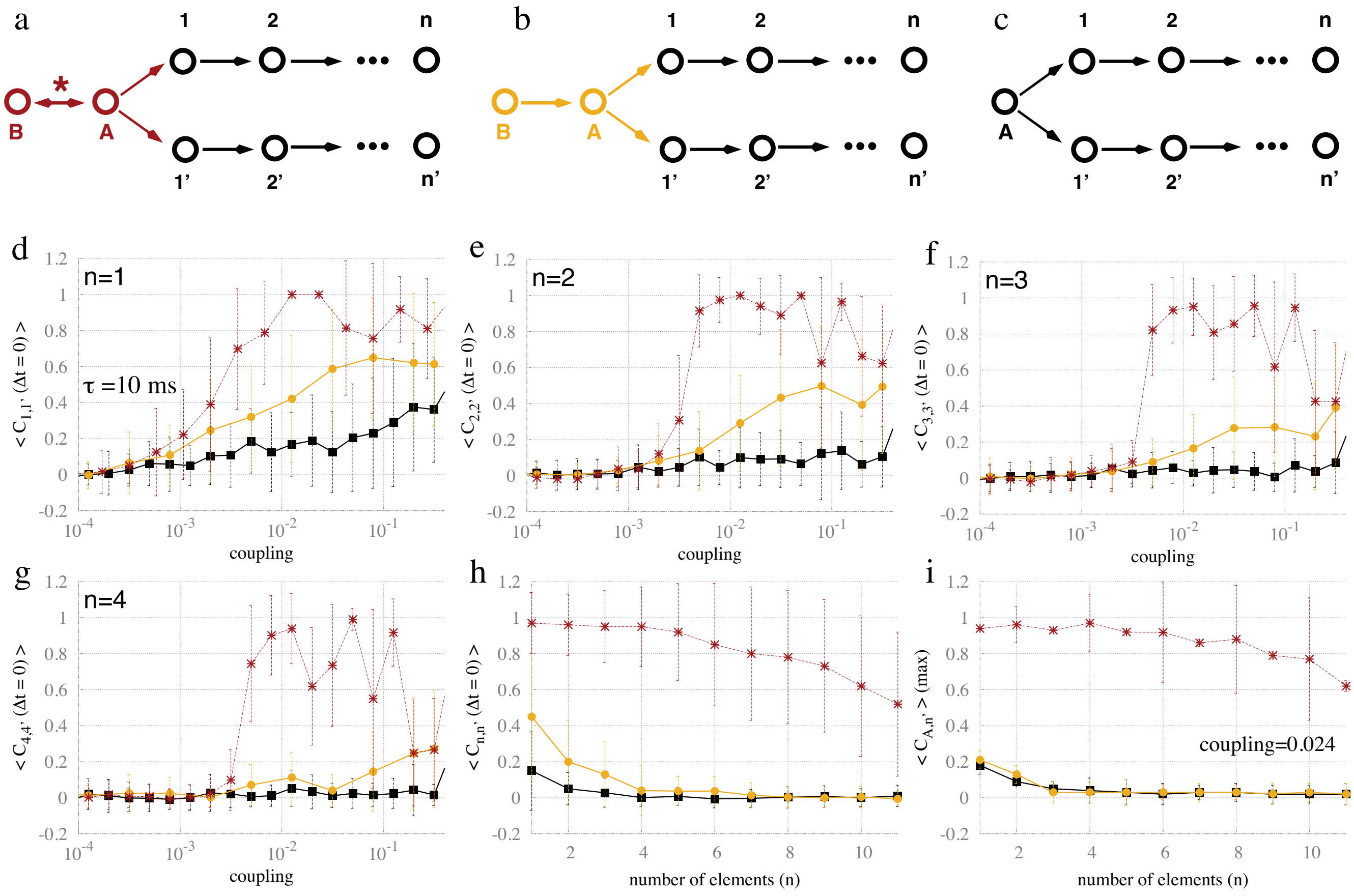

*Figure 12: Propagation of synchrony to pairs of nodes at higher orders of distance. Common driving to first (1,1'), second (2,2') and n-th (n,n') order for the resonance-induced pair (a), a unidirectional input (b), and simple common driving (c). (d) to (g): Zero-lag crosscorrelation for the different types of common driving from the first to the forth order vs. the coupling strength. (h) Zero-lag crosscorrelations between pairs of nodes (n,n') as a function of the distance from the driver node A. (i) Maximum (non-zero-lag) crosscorrelations as a function of the distance from the driver node A. Red, yellow and black curves represent the crosscorrelation averaged over 40 trials for the system depicted in (a), (b) and (c) respectively.*

## Discussion

Zero-lag synchronization between distant neuronal populations confers a number of important computational advantages, and finds broad empirical support. Here we report that common driving of passive nodes by a central "master" (motif M3), a scenario that is broadly assumed to underlie zero-lag synchrony, fails completely in the weak-coupling regime and is sensitive to parameter mismatch. However, the addition of one or more mutually coupled pairs fosters the emergence of zero-lag synchrony in the outer nodes of triplet motifs, and beyond. We find that this effect is robust to many of the particular details of the system, the spatial scale and parameter asymmetry, and can propagate through a multi-synaptic relay chain. In stark contrast, the further addition of a reciprocal connection between the driven nodes introduces frustration for delays that favor out-of-phase synchrony and fails to promote zero-lag synchronization. The disruptive effect of adding new edges that close the motif reinforces the observation that it is the topology (not the total amount of coupling) that determines the zero-lag synchrony. This is also evident by the fact that that an increase in the coupling over two orders of magnitude in the unidirectional motif (M3) is less effective than adding a single feedback connection (where the effective coupling within that pair is simply doubled).

We have denoted this reciprocal pair a *resonance pair* because it can induce zero-lag synchronization between outer nodes, principally by decreasing the irregularity of the common driving node. We find that an entire family of three- and four-node motifs exhibits zero-lag synchronization in the presence of such a resonance pair. Perhaps the archetypal motif in this family is M9 (see Fig. 1) also known as the dynamical relaying motif [24-27, 33-38, 49]. This motif contains two active resonance pairs (Fig. 1). Here we find that one feedback connection to the driver node can be removed (i.e., transforming the motif into M6) without compromising the synchronization between the outer nodes (confirming a recent observation in electronic circuits [50]). Similarly, the addition of one extra node mutually connected to the driver node, M3+1 (thereby comprising a resonance pair) causes robust zero-lag synchronization of the driven nodes where M3 alone fails. This indicates that a necessary condition for nodes 1 and 3 to synchronize is that the resonance-pair nodes also synchronize, regardless of their exact phase relationship. The synchronization of the resonance pair appears in turn to enhance its propensity to synchronize the driven nodes because when the driving node is synchronized its internal incoherence diminishes: This change in the regularity of the master node in turn enslaves the unilaterally driven node onto the synchronization manifold (Fig. 9). Thereby, we propose that the mechanism that promotes zero-lag synchronization in the dynamical relaying motif is indeed the resonance pair, in common to all other motifs in the broader family we examined.

We observed the effect of the resonance pair in a variety of different models (Hodgkin-Huxley neurons, populations of Izhikevich neurons, and neural mass models) and scales: motifs of neurons and motifs of cortical regions. The results are also robust with respect to the delay, the coupling strength, the oscillatory frequency band, and arise in autonomous, chaotic systems as well as noise-driven excitable dynamics. It seems reasonable to propose that resonance-induced synchronization will prove important for other neuronal systems, such as dendritic oscillations in single-neuron dynamics [51], and indeed other physical and biological systems of any domain characterised by weak interactions. Although the responses of neural populations to noisy inputs have been well studied [52], it remains to be seen if our results prove robust to further physiological details, including embedding stronger synaptic inputs into the noisy background [53] and stronger balanced background inhibitory and excitatory inputs [30]. We also note that although our study focused mainly on interactions with time delay, the resonance-induced synchronization can also occur in systems with no time delay (Figs. 7 b and c, and supplementary Figs. S3, S5, S7, S8 and S10).

Despite the robustness of the present effect in different classes of models and dynamical regimes, the universality and extent of the phenomenon remains to be clarified. Phase-resetting curves (PRCs) can be useful to predict whether phase or out-of-phase synchronization will arise [54]: This is a crucial factor in the dynamics because frustration does not occur in the case of in-phase synchronization. While usually studied in systems without delay, PRCs can also be used in systems in the presence of conduction delays [25, 55]. Analysis of the PRC can also be employed for formal stability analysis of synchronization of motif dynamics [25]. A second caveat, at least in the model of population of spiking neurons, is the type of dynamics studied - namely that in the dynamical regime studied here, neurons spike at least once per population cycle. An alternative approach would be to analyze synchronization in motifs of populations of spiking neurons in a sparsely synchronized regime [56] – that is when individual neurons spike less often than the background ensemble cycle. Further analysis is hence required to elucidate the extent to which our results translate to other physical and biological systems, perhaps focusing on canonical models that are more amenable to mathematical analysis such as the Kuramoto system.

Computational studies of anatomically derived brain networks have shown that motifs M9 and M6 are the first and second most abundant of all three-node motifs in the macaque visual cortex [9] and are among the most frequent motifs in other cortical networks (Fig. 1). Moreover, they appear to be clustered around the core "rich club" backbone of the structural connectome [57]. The presence of a resonant-pair in these motifs, and the robust zero-lag synchrony that they confer, may provide a dynamical advantage for these pairs. However, given the additional wiring cost, it is not clear why motif M9 is more common than M6. A possible explanation we provide derives from our observation that synchronization on motif M9 is robust to longer delays in one branch of the resonance pair in comparison to M6 (Fig. 6). Hence, the gain in robustness might overcome the cost of maintaining this extra feedback connection.

The influence of a resonance pair is not limited to local synchronization dynamics but also, through propagation, to larger networks, decaying only slowly with the polysynaptic distance (see Figs. 8 and 14). In a sufficiently sparse network like the brain, the number of neurons grows roughly exponentially with the inter-node distance. The coexistence of the slow decay (long correlation length) of the influence of the resonance pair, with rapid growth in the number of affected elements as a function of synaptic distance suggests that the zero-lag synchronization arising locally through a resonance pair has the capability to impact globally on network dynamics. Reframed in terms of a branching process, the slow decay of zero-lag synchronization and rapid growth of neuronal connectivity could lead to critical or supercritical propagation of zero-lag synchrony, consistent with prior theoretical considerations [58], and also suggesting a means for analytic extension of the present results.

The notion of motifs as fundamental building blocks of complex networks has yielded considerable prior success [9,10,53]. Degree distribution, the relative density of reciprocal synapses, convergence, divergence, and chains of synapses have been shown to play a crucial role in shaping the dynamics and synchronization properties of large networks [59-62]. In contrast to these studies, which focus on the global statistical features of large-scale networks, we have focused on particular features of small motifs. Future work, aimed at immersing these small motifs into larger networks, and focusing on the role of reciprocal nodes on the global synchronization properties of such networks, would be of significant interest. Our work confirms that the interplay between structural, functional and effective connectivity, while likely complex [63], may nonetheless be reliant upon a small number of unifying principles.

## Methods

We simulated neuronal motif dynamics at different scales, and for different dynamical scenarios. First, representing the microscopic scale, each node was taken as a single neuron. For this endeavor we utilized the Hodgkin-Huxley model. Second, at the circuit scale, we took each node as a large population of spiking neurons. Third, at the mesoscopic scale, we considered a simplified coarse-grained version in which each population was taken as a neural mass model.

**Hodgkin-Huxley neurons**

Each node was modeled by the well-known Hodgkin-Huxley equations [39]. The dynamics of the membrane potential depends on sodium, potassium, leaky, and synaptic (intra-motif and external) current components,

$$C\frac{dV}{dt} = -g_{Na}m^3h(V-E_{Na}) - g_K n^4(V-E_K) - g_L(V-E_L) + I_{syn} + I_{ext}, \quad (1)$$

where C = 1 μF/cm$^2$ is the membrane capacitance. The maximal conductances of the channels occur for completely open channels, with conductances given by $g_{Na}$ = 120 mS/ cm$^2$, $g_K$ = 36 mS/ cm$^2$, and $g_L$ = 0.3 mS/ cm$^2$. $E_{Na}$ = 115 mV, $E_K$ = −12 mV, and $E_L$ = 10.6 mV stand for the corresponding reversal potentials. Generally, the voltage-gated ionic channels are not fully opened. The probability of finding them open depends on the gating variables. The Na+ channel depends on the combined effect of gating variables m(t) and h(t), whereas K+ depends on n(t). They evolve according to the equations,

$$\frac{d\mathrm{m}}{dt} = \alpha_{\mathrm{m}}(\mathrm{V})(1-\mathrm{m}) - \beta_{\mathrm{m}}(\mathrm{V})\,\mathrm{m},$$
$$\frac{d\mathrm{h}}{dt} = \alpha_{\mathrm{h}}(\mathrm{V})(1-\mathrm{h}) - \beta_{\mathrm{h}}(\mathrm{V})\,\mathrm{h}, \quad (2)$$
$$\frac{d\mathrm{n}}{dt} = \alpha_{\mathrm{n}}(\mathrm{V})(1-\mathrm{n}) - \beta_{\mathrm{n}}(\mathrm{V})\,\mathrm{n}.$$

Hodgkin and Huxley set the empirical functions α and β to fit the experimental data of the squid giant axon,

$$\alpha_{\mathrm{m}}(\mathrm{V}) = \frac{2.5 - \mathrm{V}/10}{\exp(2.5 - \mathrm{V}/10) - 1},$$
$$\beta_{\mathrm{m}}(\mathrm{V}) = 4\exp(-\mathrm{V}/18),$$
$$\alpha_{\mathrm{h}}(\mathrm{V}) = 0.07\exp(-\mathrm{V}/20),$$
$$\beta_{\mathrm{h}}(\mathrm{V}) = \frac{1}{\exp(3 - \mathrm{V}/10) + 1}, \quad (3)$$
$$\alpha_{\mathrm{n}}(\mathrm{V}) = \frac{0.1 - \mathrm{V}/100}{\exp(1 - \mathrm{V}/10) - 1},$$
$$\beta_{\mathrm{n}}(\mathrm{V}) = 0.125\exp(-\mathrm{V}/80).$$

The synaptic current due to the interactions between neurons of the motifs are given by,

$$\tau_{syn}\frac{dI_{syn}}{dt} = -I_{syn} + \tau_{syn}\, j_e \sum_k \delta(t - t_k - \tau_k), \quad (4)$$

where $\tau_{syn}$ = 0.4 ms, $j_e = 50\ \mu A/cm^2$, and δ stands for the Dirac delta function. The summation over k stands for the spikes of the presynaptic neurons (all excitatory). $t_k$ is the time at which the k − th spike occurred. We varied the conduction delay $\tau_k = \tau$. In agreement with the literature [40], the delay τ can shape the synchronization (Fig. 3). The external current incoming to each neuron is,

$$I_{ext} = \tau_{ext}\ j_{ext} \sum_j \delta(t - t_j), \quad (5)$$

where $j$ runs over 1000 external neurons, $j_{ext} = 20\ \mu A/cm^2$, $\tau_{ext} = 1$ ms, and $t_j$ corresponds to the spike times, modeled by an independent Poisson process for each neuron with rate r = 40 Hz. As shown in supplementary Fig. S11, nearly identical results can be also obtained by assuming the external current term as synaptic contribution and including it as an extra term in equation (4) with $j_{ext} = 50\ \mu A/cm^2 j_{ext} = 50 \mu A/cm^2$. The equations were integrated by the Runge-Kutta method of fourth order, with time steps of 0.01 ms. Initial transient dynamics were discarded.

**Populations of Izhikevich neurons**

For this large-scale circuit model, each node represented populations of 500 randomly connected neurons described by the Izhikevich model [43]. 400 neurons were excitatory and 100 neurons were inhibitory. The neurons were described by the following equations:

$$\frac{dv}{dt} = 0.04\, v^2 + 5v + 140 - u + I_{syn}\,,$$

$$\frac{du}{dt} = a\,(bv - u)\,, \quad (6)$$

where v represents the membrane potential, u represents the recovery variable, accounting for the $K^+$ and $Na^+$ ionic currents, and $I_{syn}$ is the total synaptic current. The neurons have a threshold at 30 mV. Once this value is reached, v is reset to c and u to u + d. Following [44], we added dispersion to these four parameters (a, b, c and d) to account for neuronal heterogeneity. Excitatory neurons have (a, b) = (0.02, 0.2), and (c, d) = (−65, 8)+(15, −6) $\sigma^2$, where σ is a random number drawn from a uniform distribution in the interval [0,1]. Inhibitory neurons have (a, b) = (0.02, 0.2) + (0.08, −0.05) σ, and (c, d) = (-65, 2).

Each neuron receives input from 80 neurons of the same population and from 25 excitatory neurons of each afferent population. The synaptic current is given by

$$I_{syn} = -v\, g_{AMPA}(t) - (65 + v)\, g_{GABA}(t) \;, \tag{7}$$

where the dynamics of the excitatory and inhibitory synapses are described by

$$\begin{aligned} \tau_{AMPA} \frac{dg_{AMPA}}{dt} &= -g_{AMPA} + 0.5 \sum_k \delta(t - t_k - \tau_k) \;, \\ \tau_{GABA} \frac{dg_{GABA}}{dt} &= -g_{GABA} + 0.5 \sum_l \delta(t - t_l) \;, \end{aligned} \tag{8}$$

δ in the equations above stands for the Dirac delta function. The summation over k (l) stands for the spikes of the presynaptic excitatory (inhibitory) neurons. $t_k$ ($t_l$) is the time at which the k − th excitatory (or l − th inhibitory) spike occurred. Conduction delays $\tau_k = \tau$, associated with excitatory long-range connections, varied. We modeled short-range (intra-node) connections with negligible delays. Synapses were modeled by exponential decay functions [64], with time constants $\tau_{AMPA}$ = 5.26 ms for excitatory and $\tau_{GABA}$ = 5.6 ms for inhibitory synapses. Each neuron was subject to an external driving given by independent Poisson spike trains, resulting from 100 excitatory neurons, at a rate of r = 16 Hz, which was also included in the sum over excitatory postsynaptic contributions (k index) of the equations above. With these parameters, individual neurons fire spontaneously, although not periodically.

The equations were integrated using a fixed-step first-order Euler method with time steps of 0.05 ms, starting with random initial conditions. To avoid spurious synchronization at the onset of simulations, neural populations were activated with random noise in 600 ms sequential windows (with a 500 ms overlap). The first transients of 1 s were discarded before further analysis.

**Neural mass models**

The preceding large-scale circuit model is a high dimensional system. Whilst the dynamics are instructive, the large number of parameters and equations preclude an intuitive perspective of the system. We therefore additionally studied a reduced system [65], which represents the large cortical scale that permits characterization of the system dynamics with respect to the most salient parameters. In contrast to the previous models, the coupling is not through discrete pulses, but by means of smooth sigmoidal rate functions, which embody population-wide neuronal responses to synaptic inputs in the presence of parameter and state dispersion [66]. This also allows us to study the robustness of the resonance-induced synchronization in relationship to the precise details - and dynamical regime - of the models.

Each node represents the mean dynamics of an ensemble of neurons, with spontaneous dynamics arising from the interaction between excitatory and the inhibitory sub-populations. The model is derived from the biophysical Morris-Lecar model [45], extended to a neural mass model with passive diffusive chemical [46], then synaptic interactions [6] and subsequently extended to large networks to model whole brain activity [5]. We utilize this most recent approach developed by Honey et al. [5,47] systematically varying the features of the connectivity: architecture, coupling strength, and delay.

This neural mass model comprises three state variables: The mean membrane potential of the excitatory pyramidal neurons, V; the mean membrane potential of the inhibitory interneurons, Z; and the average number of open potassium ion channels, W. Our main focus is on the dynamics of the pyramidal neurons. Their average membrane potential V depends on the passive leak conductance, and on the conductance of voltage-gated channels of sodium, potassium and calcium ions. The flow of current across the local pyramidal cell membranes, assumed as capacitors, governs its dynamics. In turn, the local activity of the inhibitory interneurons is course-grained modeled; its dynamics is modulated by the activity of the pyramidal cell. For each ensemble *i*, the equations for the dynamics of the mean membrane potential of the neurons are given by

$$\begin{aligned}\frac{dV^i(t)}{dt} = & -\{g_{Ca} + r_{NMDA}\, a_{ee}[(1-c)Q_V^i + c\langle Q_V^j(t-\tau)\rangle]\}\, m_{Ca}\big(V^i(t) - V_{Ca}\big) \\ & -\{g_{Na}m_{Na} + a_{ee}[(1-c)Q_V^i + c\langle Q_V^j(t-\tau)\rangle]\}\big(V^i(t) - V_{Na}\big) \\ & -g_K\, W^i(t)\big(V^i(t) - V_K\big) - g_L\big(V^i(t) - V_L\big) - a_{ie}Z^i(t)Q_Z^i + a_{ne}\, I_\delta\,; \end{aligned} \tag{9}$$

$$\frac{dZ^i(t)}{dt} = b\left(a_{ni}\, I_\delta + a_{ei}\, V(t)\, Q_V^i(t)\right). \tag{10}$$

The fraction of channels open $m_{ion}$ are the *neural-activation function*, whose shape reflects a sigmoidal-saturating grow with V

$$m_{ion} = 0.5\left[1 + \tanh\left(\frac{V^i(t) - T_{ion}}{\delta_{ion}}\right)\right] \quad . \qquad (11)$$

The third differential equation of each node *i* stands for the fraction of open potassium channels:

$$\frac{dW^i(t)}{dt} = \frac{\varphi\,[m_k - W^i(t)]}{\tau_W} \quad . \qquad (12)$$

The neuronal firing rates ($Q_V^i$, and $Q_Z^i$) averaged over the ensemble are assumed to obey Gaussian distributions, thereby giving rise to the sigmoidal activation functions [66],

$$Q_V^i = 0.5\; Q_{Vmax}\left[1 + \tanh\left(\frac{V^i(t) - V_T}{\delta_V}\right)\right] \quad ;$$

$$Q_Z^i = 0.5\; Q_{Zmax}\left[1 + \tanh\left(\frac{Z^i(t) - Z_T}{\delta_Z}\right)\right] \quad . \qquad (13)$$

Our simulations employ the previously published parameter values: $g_{Ca}$=1.1, $r_{NMDA}$=0.25, $a_{ee}$=0.4, $V_{Ca}$=1, $g_{Na}$=6.7, $V_{Na}$=0.53, $g_K$=2, $V_K$=-0.7, $g_L$=0.5, $V_L$=-0.5, $a_{ie}$=2, $a_{ne}$=1, $I_\delta$=0.3, b=0.1, $a_{ni}$=0.4, $a_{ei}$=2, $T_{Ca}$=-0.01, $T_{Na}$=0.3, $T_K$=0, $\delta_{Ca}$=0.15, $\delta_{Na}$=0.15, $\delta_K$=0.3, φ=0.7, $\tau_W$=1, $Q_{Vmax}$=1, $V_T$=0, $\delta_V$=0.65, $Q_{Zmax}$=1, $Z_T$=0, and $\delta_Z$=0.65 were set to physiological values taken from [6]. These are associated with aperiodic fluctuations arising without external noise, but rather due to homoclinic chaos [6]. Equation 9 includes the other important parameters in our analysis: $j = 1, ..., N$, the presynaptic neighboring (afferent) regions of region *i*; c, the coupling strength between cortical regions; τ, the synaptic delay between cortical regions. The model was simulated in Matlab (Math Works) at a time resolution of 0.2 milliseconds using the function *dde23*.

## Supplementary Figures

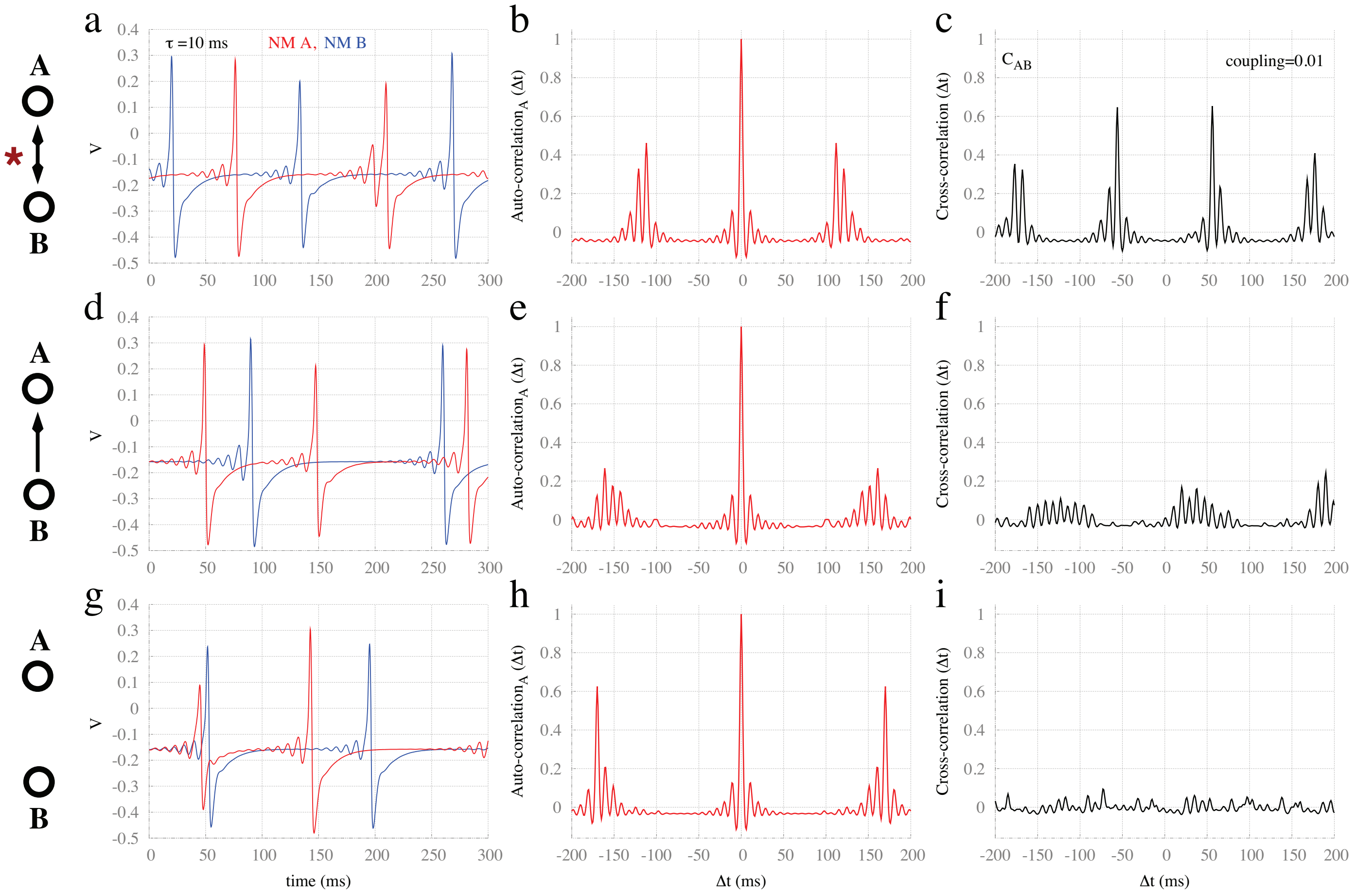

*Figure S1: Dynamics of pairs of neural mass models. (a), (d) and (g) show the time traces of the average membrane potential of the excitatory pyramidal neurons; (b), (e) and (h) show the auto-correlation function of node A; (c), (f) and (i) show the cross-correlation function between nodes A and B; respectively for a pair bidirectionally connected, unidirectionally connected, and disconnected nodes. Parameters are c = 0.01, and τ = 10 ms.*

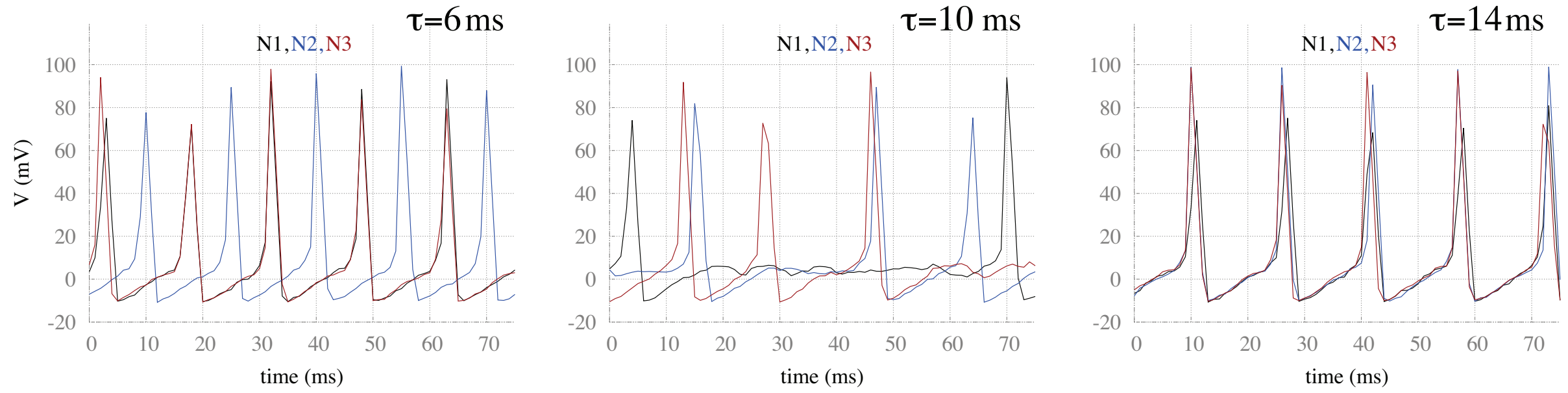

*Figure S2: Example dynamics of Hodgkin-Huxley neurons coupled on motif M6 for different time delays. From left to right, panels show anti-phase synchronization, no synchronization, and phase synchronization for increasing time delays.*

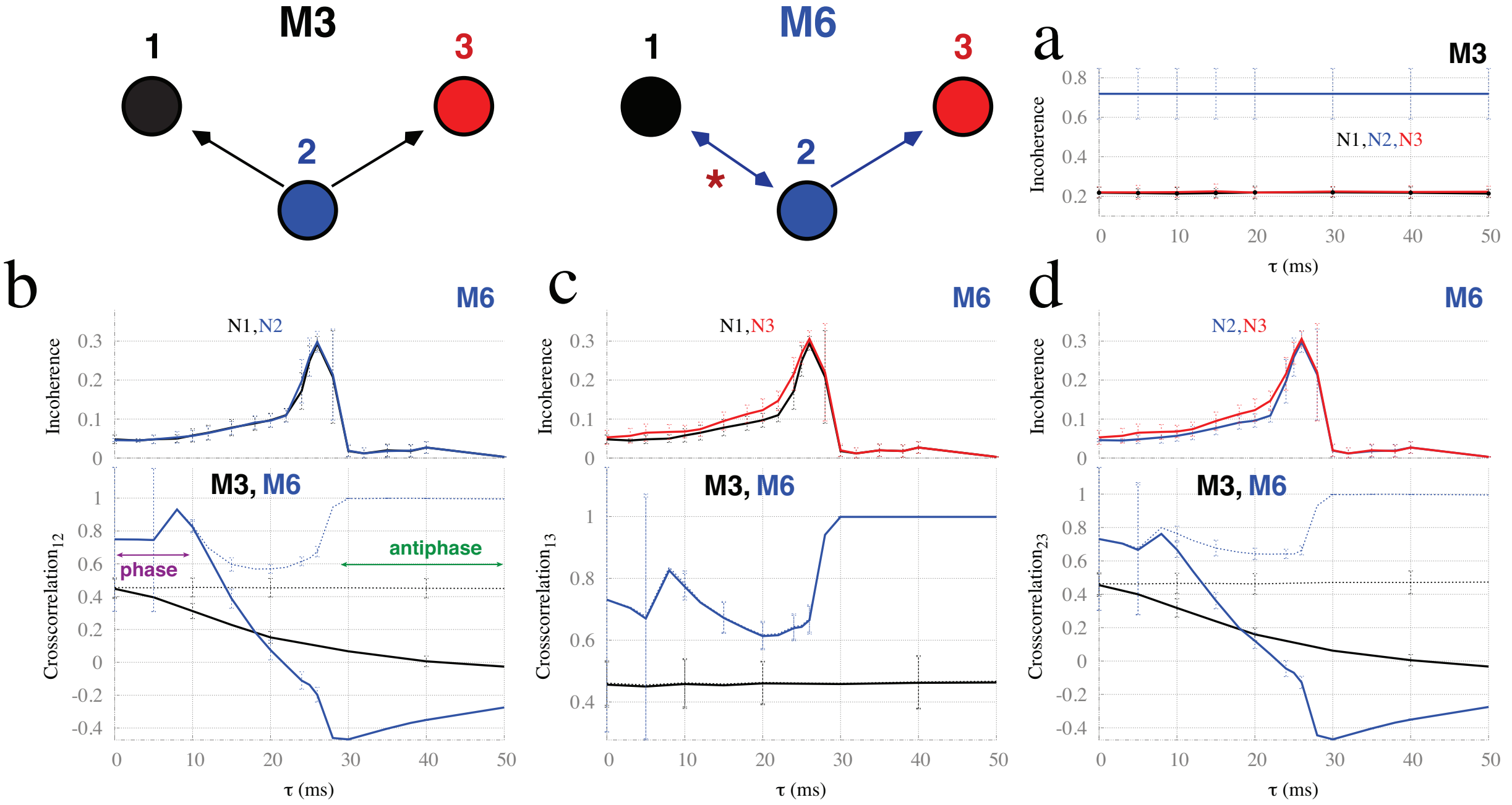

*Figure S3: Synchronization dynamics and incoherence in populations of Izhikevich neurons. Panels (a-d) as per Fig. 3 but for populations of spiking neurons. Phase, anti-phase synchrony, and a state of phase synchrony at the slow rhythm and anti-phase synchrony at the fast rhythm can be found in motif M6 depending on the time delay (see exemplar time traces in supplementary Fig. S4).*

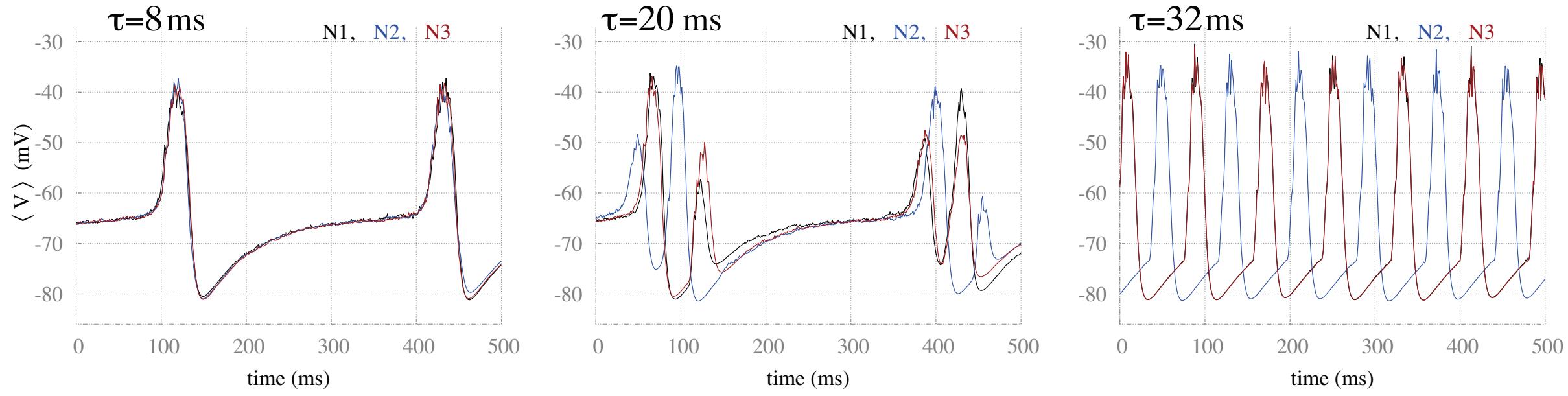

*Figure S4: Example dynamics of populations of Izhikevich neurons coupled as motif M6 for different time delays. From left to right, panels show phase synchronization, phase synchronization at the slow rhythm and anti-phase synchronization at the fast rhythm, and anti-phase synchronization respectively.*

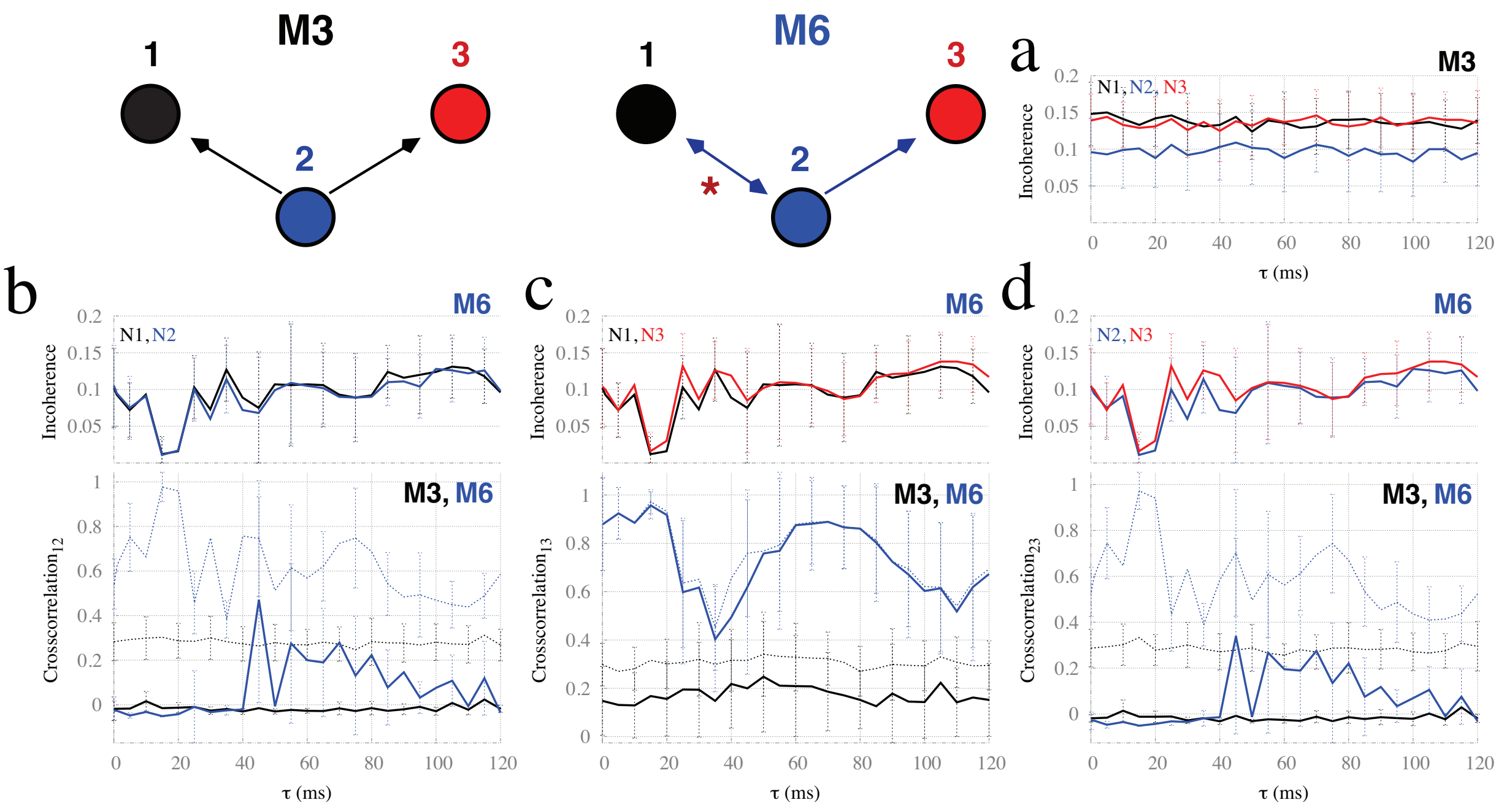

*Figure S5: Synchronization dynamics and incoherence in weakly coupled neural mass models. Panels (a-d) as per Fig. 3 but for neural mass models with coupling strength c = 0.01. Anti-phase synchrony at the slow or at the fast rhythms, and a state of low synchrony can be found in motif M6 depending on the time delay (see exemplar time traces in supplementary Fig. S6).*

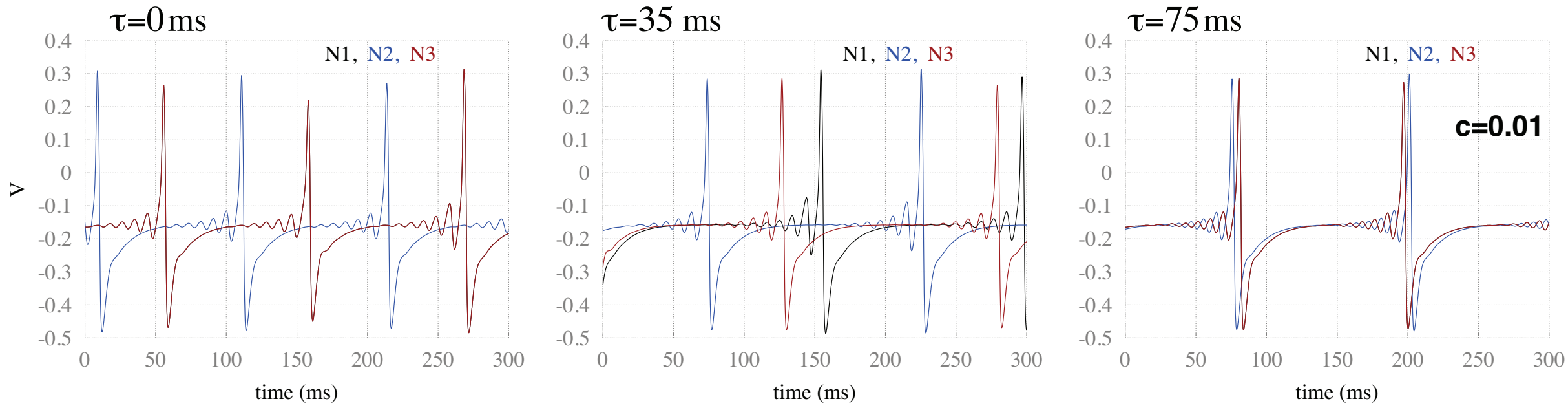

*Figure S6: Example dynamics of neural mass models coupled on motif M6. From left to right, panels show anti-phase synchrony at the slow rhythm (no time delay), weak synchrony (τ=35 ms), and anti-phase synchrony at the fast timescale (τ= 75ms). The coupling strength is weak, c = 0.01.*

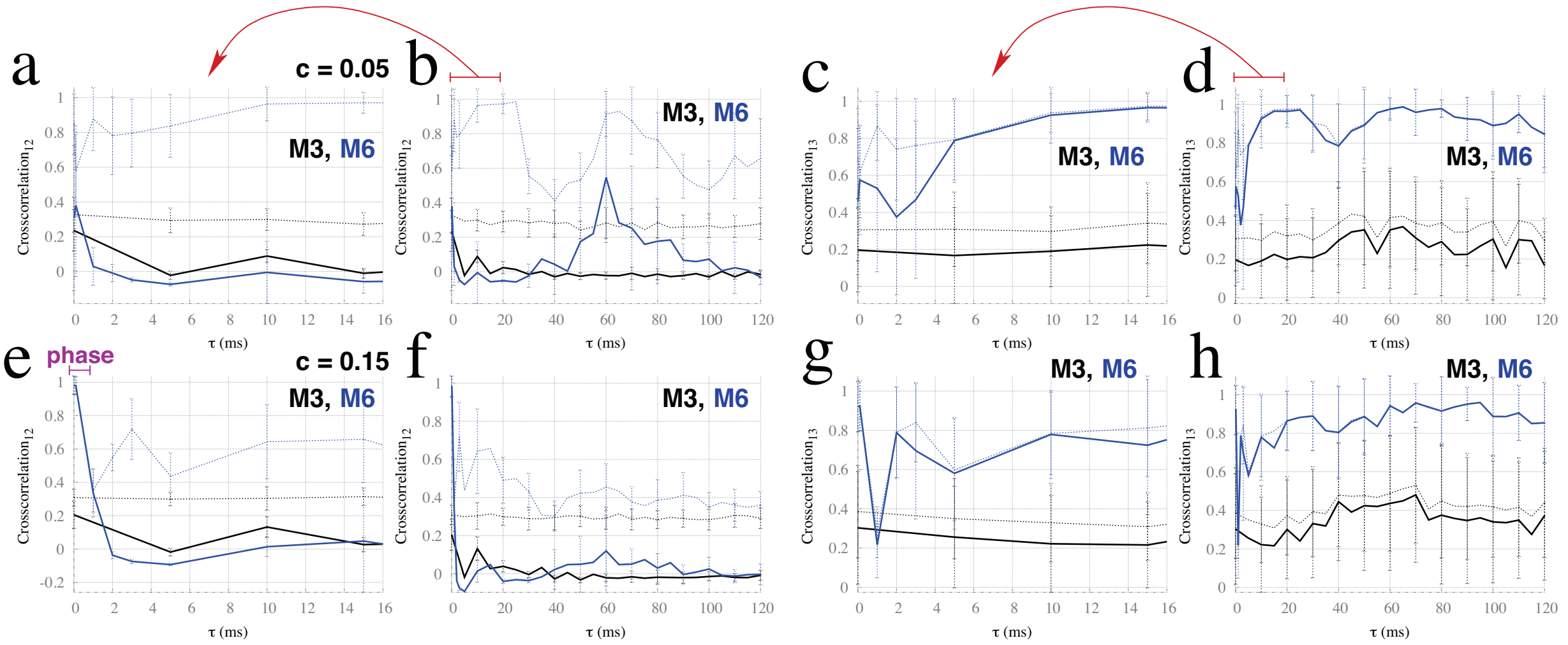

*Figure S7: Cross-correlation for strongly coupled neural mass models. Top panels show the cross-correlations between nodes 1 and 2 (a-b), and nodes 1 and 3 (c-d) as a function of the delay for coupling strength c = 0.05. Bottom panels panels show the cross-correlations between nodes 1 and 2 (e-f), and nodes 1 and 3 (g-h) as a function of the delay for coupling strength c = 0.15. First and third columns correspond to a zoom of second and forth columns respectively. Black (blue) lines represent results for motif M3 (M6), and continuous (dashed) lines represent the crosscorrelation at zero lag (maximum for all time lags). Phase synchrony, and complex synchronous states can be found in motif M6 depending on the time delay τ (see exemplar time traces in supplementary Fig. S8). Results are averaged over 40 trials.*

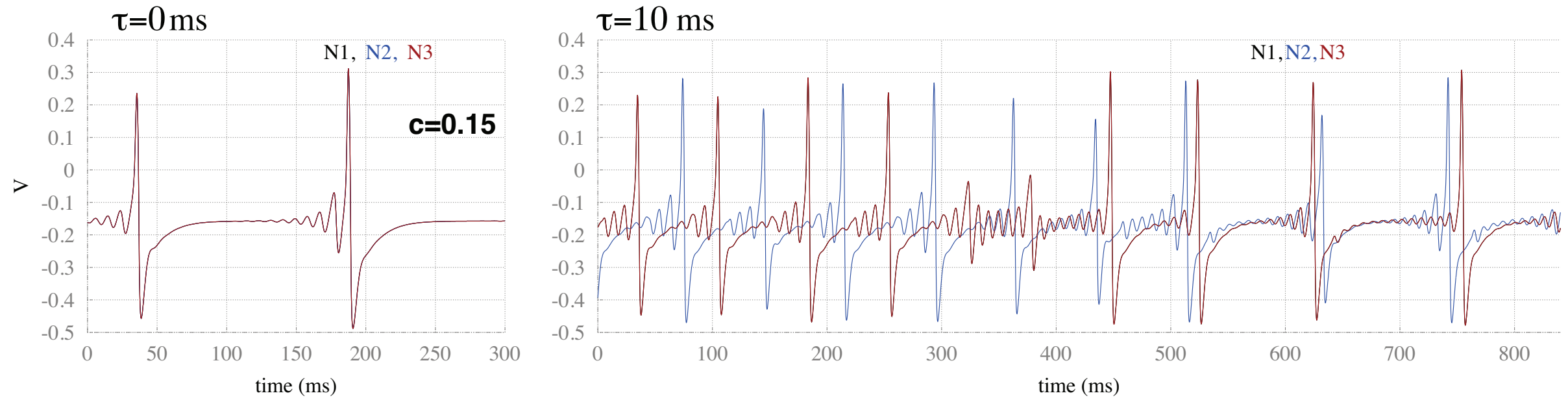

*Figure S8: Example dynamics of neural mass models strongly coupled as motif M6 for different time delays. From left to right, panels show phase synchrony, and a transition from a state of anti-phase synchrony at the slow rhythm to a state of out-of-phase synchrony. The coupling strength is c = 0.15.*

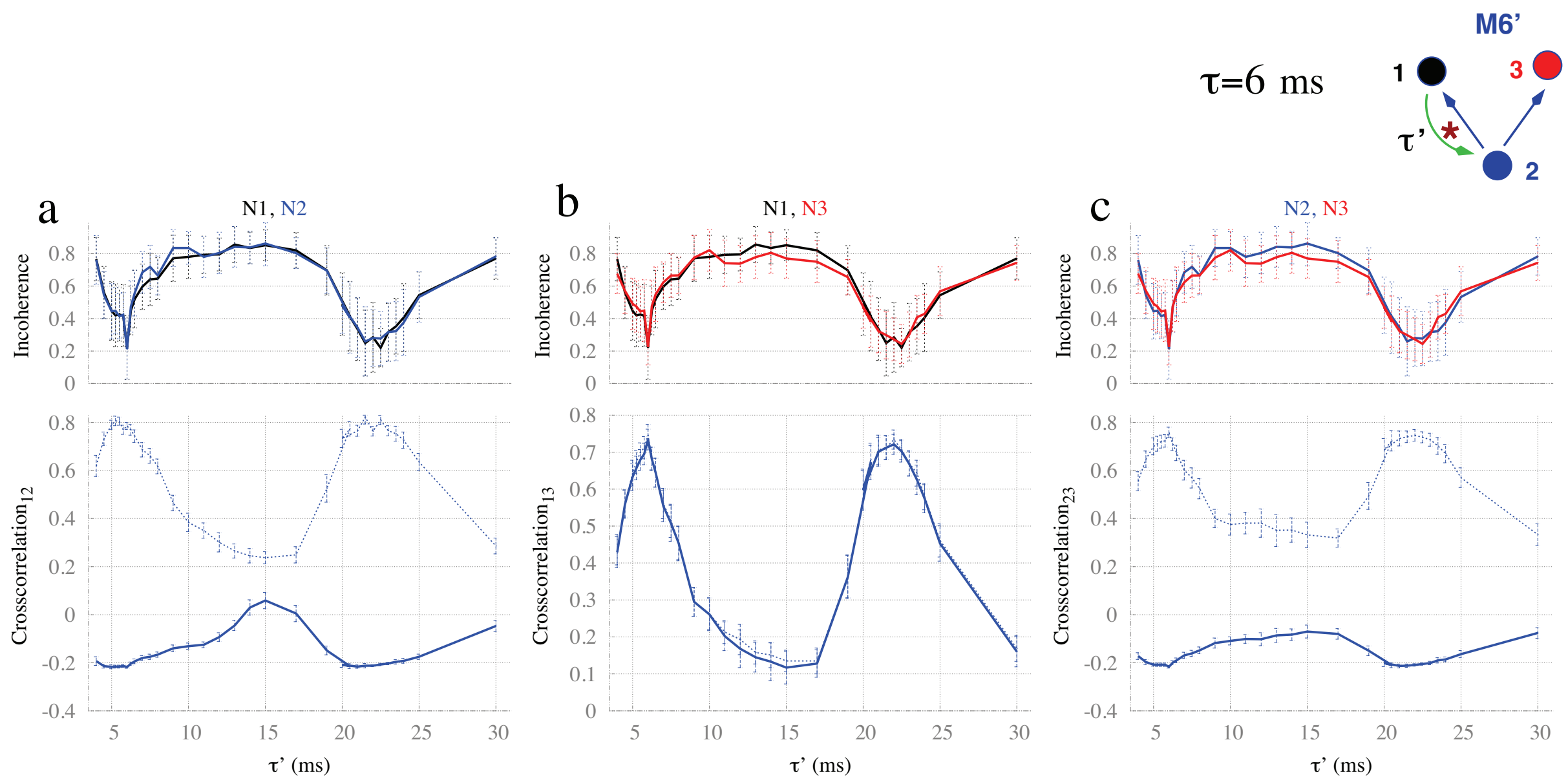

*Figure S9: Reciprocal connections give rise to incoherence and synchronization depending on the time delay mismatch in Hodgkin-Huxley neurons. (a-c) Top panels show incoherence: Colors represent different nodes. Bottom panels show cross-correlations for motif M6'. Continuous lines indicate the cross-correlation coefficients at zero time lag, and dashed lines indicate the maximum cross-correlation coefficients across all time lags. Panels a, b and c represent pairs of nodes: 1-2, 1-3, and 2-3 respectively. Results are averaged over 40 trials.*

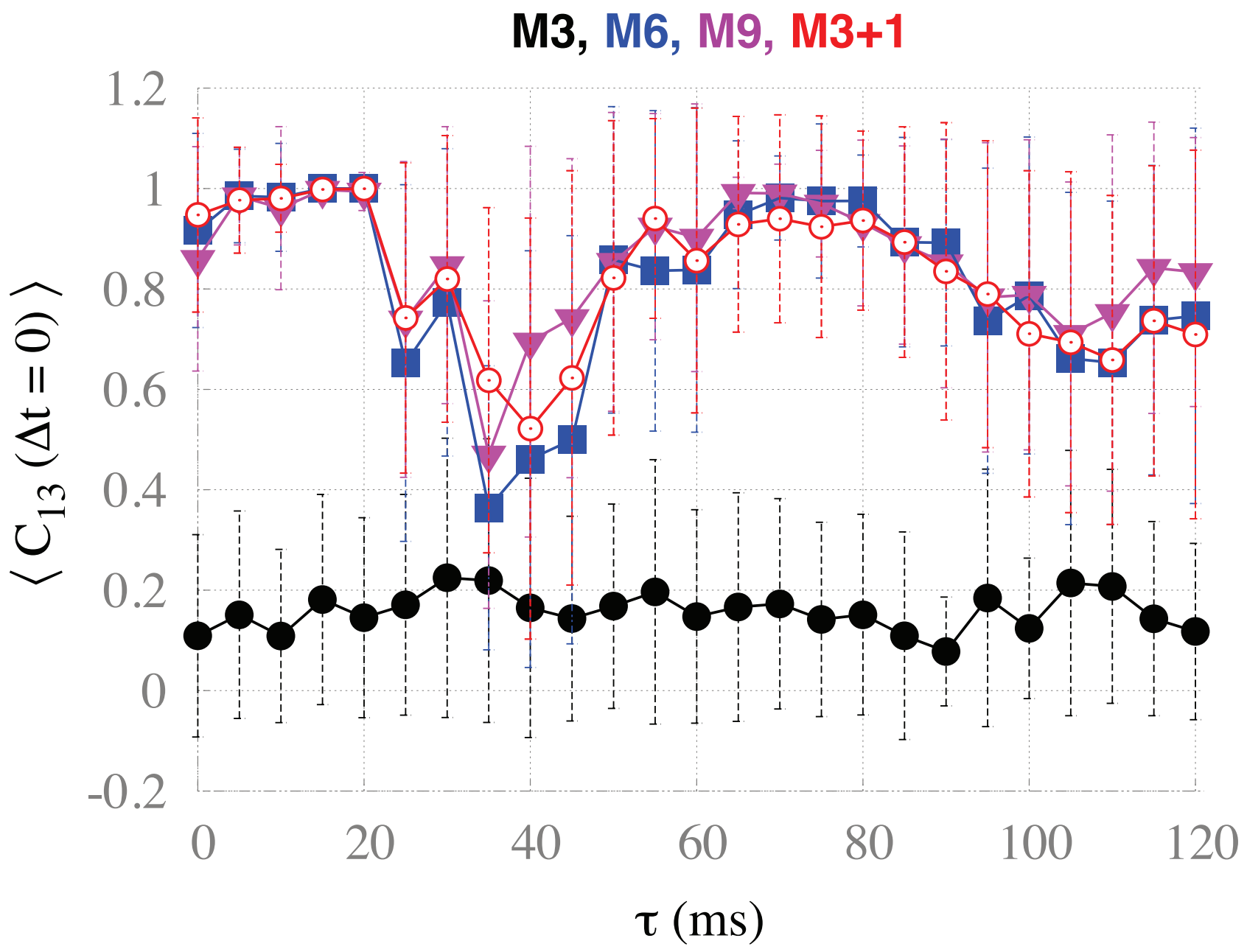

*Figure S10: Zero-lag synchronization dependence on the delay τ in the motifs of neural mass models. In agreement with [39] the synchronization depends on the coupling delay for long delays. The*

*coupling strength is c = 0.01. Crosscorrelation is averaged over 40 trials.*

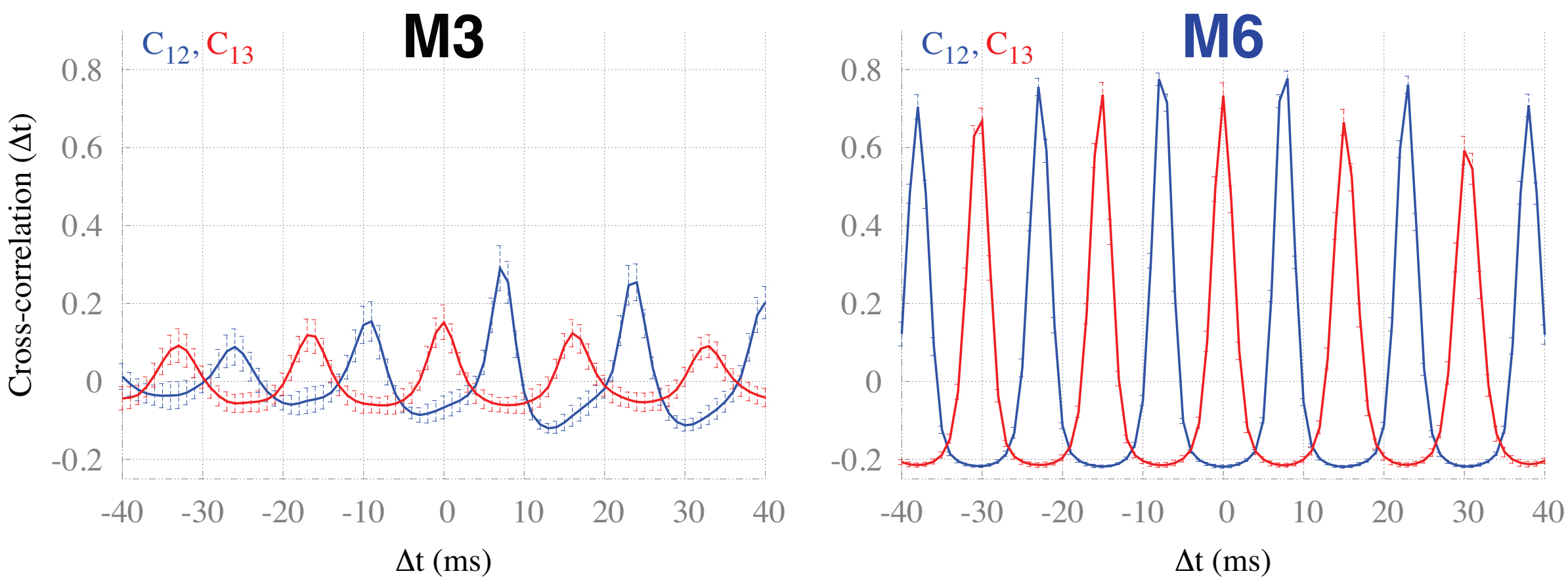

*Figure S11: Same kernel test for external driving of the Hodgkin-Huxley neurons. Nearly identical cross-correlation functions appear when the external driving is considered identical to the spikes within the motifs (see Fig. 2, panels d and h). Plot corresponds to an average over 40 trials.*